\documentclass{aa}  

\def\xmm{XMM-{\it Newton}}
\def\swift{{\it Swift}}
\def\chandra{{\it Chandra}}
\def\asca{{\it ASCA}}

\usepackage{graphicx}
\usepackage{color}
\usepackage{natbib}
\usepackage{amsmath}
\usepackage{amssymb}
\usepackage{booktabs}
\bibpunct{(}{)}{;}{a}{}{,}
\usepackage{txfonts}
\begin{document} 
  \title{Short time-scale X-ray spectral variability in the Seyfert 1 galaxy NGC 3783}

   \subtitle{}
\authorrunning{D. Costanzo\inst{1,2} et al.}
   \author{D. Costanzo\inst{1,2}, M. Dadina\inst{2}, C. Vignali\inst{1,2}, B. De Marco\inst{3}, M. Cappi\inst{2}, P. O. Petrucci\inst{4}, S. Bianchi\inst{5}, G. A. Kriss\inst{6}, J. S. Kaastra\inst{7,8}, M. Mehdipour\inst{6,7}, E. Behar\inst{9}, G. A. Matzeu\inst{1,2} 
          }

   \institute{Dipartimento di Fisica e Astronomia "Augusto Righi", Università degli Studi di Bologna, Via Gobetti 93/2,
40129 Bologna, Italy\\ \email{deborah.costanzo2@unibo.it} \and INAF – Osservatorio di Astrofisica e Scienza dello Spazio di Bologna, Via Gobetti 93/3, 40129 Bologna, Italy \and Departament de Física, EEBE, Universitat Politecnica de Catalunya, Av. Eduard Maristany 16, E-08019 Barcelona, Spain \and Univ. Grenoble Alpes, CNRS, IPAG, 38000 Grenoble, France\and Dipartimento di Matematica e Fisica, Università degli Studi Roma Tre, via della Vasca Navale 84, 00146 Roma, Italy \and Space Telescope Science Institute, 3700 San Martin Drive, Baltimore, MD 21218, USA\and SRON Netherlands Institute for Space Research, Niels Bohrweg 4, 2333 CA Leiden, The Netherlands\and  Leiden Observatory, Leiden University, P.O. Box 9513, 2300 RA Leiden, The Netherlands\and Department of Physics, Technion, Haifa 32000, Israel
               }

   \date{Received xxx; accepted xxx}

 
  \abstract
    {}
   {We report on the X-ray time resolved spectral analysis of XMM-Newton observations of NGC 3783. The main goal is to detect transient features in the Fe K line complex, in order to study the dynamics of the innermost accretion flow.}
   {We reanalize archival observations of NGC 3783, a bright local AGN, for which a transient Fe line was reported, complementing this data set with new available observations. This results in a long set of observations which can allow us to better assess the significance of transient features and possibly test their recurrence time. Moreover, since the new data catch the source in an obscured state, this analysis allows also to test whether the appearance/disappearance of transient features is linked to the presence of obscuring gas.}
   {We detect discrete features at the $\geq90\%$ significance level both in emission and in absorption at different times of the observations, split into 5ks time-resolved spectra. The overall significance of individual features is higher in the obscured dataset. The energy distribution of the detections changes between the two states of the source, and the features appear to cluster at different energies. Counting the occurrences of emission/absorption lines at the same energies, we identify several groups of $\geq3\sigma$ detections: emission features in the 4-6 keV band are present in all observations and are most likely due to effects of the absorber present in the source; an emission line blend of neutral Fe K$\beta$/ionized Fe K$\alpha$ is present in the unobscured dataset; absorption lines produced by gas at different ouflowing velocities and ionization states show an increase in energy between the two epochs, shifting from $\sim6.6$ keV to $\sim6.7-6.9$ keV.
   The representation of the features in a time-energy plane via residual maps highlighted a possible modulation of the Fe K$\alpha$ line intensity, linked to the clumpiness of the absorbing medium.}
 {}
 

\maketitle

\section{Introduction}

Active Galactic Nuclei (AGN) show variability at any wavelength over a wide range of time-scales, from minutes to years (e.g. \citealt{padovani2017}). In the X-ray band, variability is found to be much faster than in other wavelenghths; this is ascribed to variable phenomena happening in the innermost regions of the accretion flow (corona and inner disk), i.e. at a few --tens  gravitational radii ($\mathrm{R_g=GM/c^2}$) from the supermassive black hole (SMBH) \citep[e.g][]{mushotzky1993}. Moreover, these same variable phenomena may drive (at least part of) the variability from the outer accretion disk, observed at
longer wavelengths (UV and optical) and over longer timescales \citep{kcg01,au06}. Observing diverse timescales for variability in distinct parts of the spectrum can help us to distinguish different processes in action and understand where they are possibly originating.

Seyfert 1 galaxies are the best targets to study these regions, because they are the brightest in X-rays and are thought to be observed "face-on" from an angle typically $\leq30^\circ$, i.e. with a clear and direct view of the accretion flow \citep{urrypadovani95}. Their X-ray spectra show many emission features, the most prominent being the fluorescent neutral Fe K$\alpha$ line at 6.4 keV. This line likely originates from the reflection of the primary continuum on the accretion disk. We expect that its shape and energy are altered by the proximity to the SMBH \citep{fabian2000} and/or by the geometry/physical state of the accretion flow. Absorption features can also be present, suggesting the presence of outflows (winds) along the line of sight; their dynamics and changes in their physical state (e.g., density and ionization) can be traced studying absorption-line variability \citep{pounds03,reeves03,dadina2005,tombesi2010}.
While variability patterns have been intensively studied for emission lines \citep[e.g.][]{iwasawa2004,demarco2009}, an analysis of variable absorption features is generally more challenging, but it holds the potential to unveil the location of winds. Moreover, a simultaneous study of emission and absorption variability could highlight possible correlations between accretion and ejection phenomena, that could deepen our understanding of the launching mechanism for outflows.

The goal of this paper is to carry out a systematic and simultaneous time-resolved spectral analysis of emission/absorption features in the spectra of an X-ray bright Seyfert galaxy, NGC~3783. This source presents interesting properties for the study of variable emission/absorption lines, as detailed below, and is one of brightest with long \xmm\ EPIC/pn observations.
Some of these observations, those taken in 2000 and 2001, were already analyzed with time-resolved spectral analysis \citep{tombesi2007,demarco2009}, focusing on the variability properties of the emission features. We are adding a characterization of the absorption lines present in those same datasets and furthermore we are analyzing new observations taken in 2016. This will allow us to study also the variations on long time scales of the emission/absorption lines.

NGC~3783 is a Seyfert 1 galaxy at redshift $z=0.0097$ \citep{theureau1998}, with a SMBH mass of $\mathrm{M=3.0 \pm 0.5 \times 10^{7}\ M_{\odot} }$, as estimated via reverberation mapping studies in optical and UV bands \citep{peterson2004}.
Analyzing the \xmm\ observation taken in 2001 \citep[hereafter R04]{reeves2004}  identified several features: a strong Fe K$\alpha$ at 6.4 keV, an emission line at $\sim$7 keV due to a blend of neutral Fe K$\beta$ and H-like Fe, and absorption at $\sim$6.6 keV due to highly ionized Fe, plus an absorption edge at $\sim$7.1 keV.  In the same dataset, \cite{tombesi2007} found a modulation of the flux and a correlated variation of the Fe K$\alpha$ with a broad redshifted component on time-scales of $\sim$27 ks. 
In December 2016, during a \swift/XRT monitoring program, NGC~3783 showed  heavy X-ray absorption produced by an obscuring outflowing gas \citep[M17 hereinafter]{mehdipour2017}. As a result, new absorption lines from Fe~XXV and Fe~XXVI appeared in the \xmm\ spectrum. From their analysis, M17 found for the outflow a column density of a few 10$^{23}$~cm$^{-2}$ and a velocity of few 1000 km~s$^{-1}$, and interpreted it as a clumpy, inhomogeneous medium consistent with clouds at the base of a radiatively-driven disk wind located in the outer broad line region of the AGN.
A similar but less intense obscuration event was revealed by \cite{kaastra18} in a \chandra/HETG observation taken in Aug. 2016, with a column density of one order of magnitude lower than the one derived in Dec. 2016. Using all the X-ray observations of NGC~3783, from 1993 (\asca) to 2016 (\chandra), it was clear that the source displayed an absorption column density larger than 10$^{22}$ cm$^{-2}$ in roughly 50$\%$ of the observing time. 
\citet[hereafter DM20]{demarco20} constrained the short time scales (from about one hour to ten hours) variability properties of the obscurer in the 2016 \xmm\ dataset. Their spectral-timing analysis showed that the observed fast variations in the soft X-rays were consistent with changes in the ionization parameter. This study allowed inferring a recombination time of $\lesssim$1.5 ks, corresponding to a lower limit on the electron density of n$_e\sim7.1\times10^{7}$ cm$^{-3}$. This value is consistent with M17 results and places the obscurer at a distance between 7 and 10 light days. 

Here, we present the results from the overall XMM-Newton available datasets. We present a comprehensive and detailed time-resolved spectral analysis aimed at detecting variable line-like features, either in emission or in absorption (Sect.~\ref{variablefeaturessearch}). The statistical significance of the analysis is then estimated in Sect.~\ref{simulations} by using detailed Monte Carlo simulations. Non-random patterns of variability are searched for (Sect.~\ref{residualmaps}) and overall findings are then discussed (Sect.~\ref{discussion}).

\section{XMM-Newton observations and data reduction}	
\label{Observations and data reduction}
\begin{figure*}
	\centering
	\resizebox{\hsize}{!}{
	\includegraphics{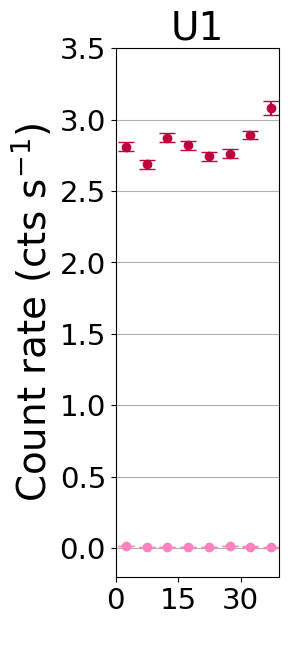}
	\includegraphics{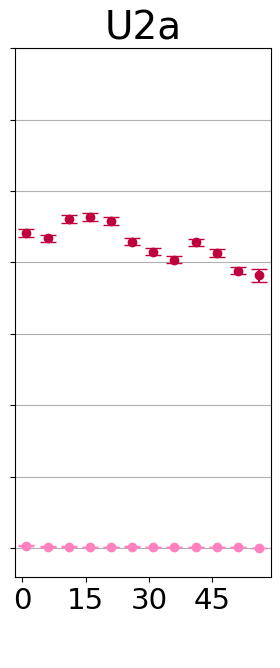}
	\includegraphics{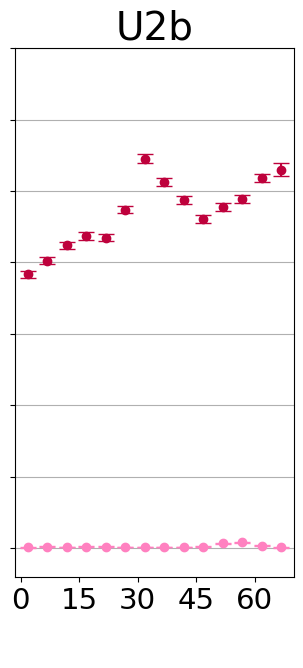}
	\includegraphics{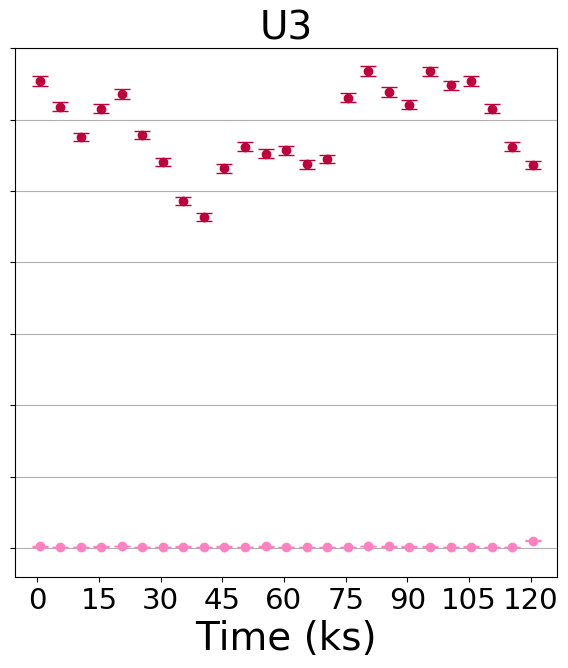}
	\includegraphics{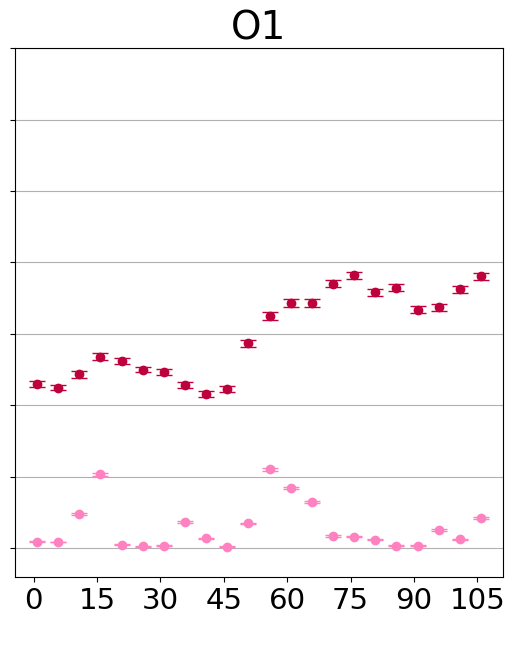}
	\includegraphics{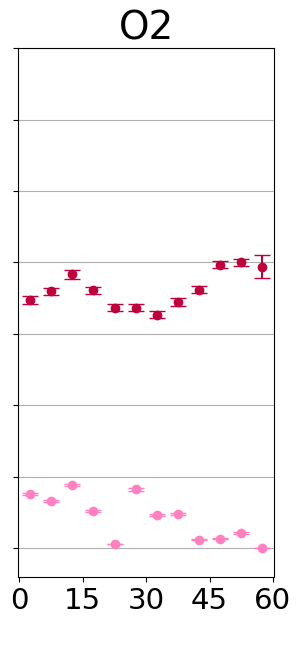}}
	\caption{Background-subtracted light curve of NGC 3783 (red points) with the corresponding background (pink points), extracted in the 4-10 keV band. The time bins are 5ks. The identifiers of each observation (U1-3, O1-2) refer to the state of the source (Unobscured, Obscured), as pointed in Sect.~\ref{Observations and data reduction}.}
	\label{lightcurves}
\end{figure*}

There are seven observations of NGC~3783 in the \xmm\ Science Archive, but two of them (OBS ID 0112210401 and OBS ID 0112210601) were excluded from our analysis because of their short duration ($\sim$ 4~ks). The remaining observations we analyzed are listed in Table~\ref{datasetngc3783}. Following M17 and DM20 we identify two epochs, corresponding to the state of the source: observations from 2000 and 2001 caught the source in an unobscured state, and are hereafter identified as U1, U2 and U3, while during the observations taken in 2016 the source was in an obscured state, so they are named as O1 and O2.
\begin{table}
	\centering                          
	\begin{tabular}{ccccc}        
		\hline
		&Obs ID  & Start date &Flux& Exposure\\ 
		&  & \tiny{DD/MM/YYYY} && \tiny{ks} \\ 
			    \hline
		U1 &0112210101&28/12/2000 &3.91 &26 \\
		U2 &0112210201&17/12/2001 &3.13 &87\\
		U3 &0112210501&19/12/2001 &4.06 &87\\
		O1 &0780860901&11/12/2016 & 2.11 &76\\
		O2 &0780861001&21/12/2016 &2.54 &38\\
        \hline
	\end{tabular}
\caption{Log of the XMM observation of NGC 3783 used in this work. The columns report: 1) the nomenclature used throughout the paper to refer to each observation; 3) the observation ID; 3) the observation date; 4) the 4.5-10 keV flux (in units of $10^{-11}$ erg s$^{-1}$ cm$^{-2}$) taken from  4XMM-DR10 catalogue \citep{webb2020}; 5) the effective exposure (after removal of flares). Note that observation U2 is split in two parts (U2a and U2b, of 40ks and 47ks respectively), separated by $\sim$4ks.}
   \label{datasetngc3783}
\end{table}
We used data from the EPIC pn detector only, because of its higher effective area with respect to the MOS. The camera was operated in Small Window mode during all observations. Data were reduced using the XMM-SAS v.16.1, following standard procedures. After running the \texttt{epproc} task, U2 resulted in being split in two parts (U2a and U2b), separated by $\sim$~4 ks. Source photons were collected from circular regions of 45 arcsec radius, while background photons were extracted from an offset circle of 65 arcsec radius, close to NGC~3783 and free of any background sources. Observations U1, U2 and U3 show limited or no high-background intervals, which were present only at the end of the pointing. Observations O1 and O2 were affected by several flares, removed by imposing a threshold in the count rate at 0.2 cts s$^{-1}$ for energies > 10 keV. The resulting exposure time of the cleaned event files are reported in Table~\ref{datasetngc3783}. The background-subtracted light curves were obtained with the SAS routine \texttt{epiclccorr} and are shown in Fig.~\ref{lightcurves}. Using the \texttt{epatplot} task, we verified the absence of pile-up. Response matrices were produced with the SAS tasks \texttt{arfgen} and \texttt{rmfgen}, and spectra were analyzed using the XSPEC v.12.9 software package \citep{arnaud96}.
 
\section{Variable features search}
\label{variablefeaturessearch}

In order to probe the closest regions to the SMBH, we need to sample the shortest possible variability time-scales. On these time scales, transient/variable emission/absorption features may manifest, for example due to the variability of the primary irradiating X-ray continuum, or to motions of the outflowing gas. 
To study these features, we relied on time-resolved spectral analysis techniques. The aim of our approach is to eventually find features that would not be detected in average spectra: in fact, because of their transient/variable nature, they could be averaged out over long exposure times. Nonetheless, there is an observational limit to the proximity to the SMBH (and minimum time-scale) which is set by the photon count-rate available. Therefore the time resolution of our spectra corresponds to a trade off between sampling short time scales while retaining a sufficient number of counts within each time-resolved spectrum to allow for a meaningful statistical analysis. In order to identify the physical scales that we can probe, we first considered the Keplerian orbital period at a given distance (in $\mathrm{R_{g}}$) from a black hole with adimensional spin $a$ \citep{bardeen1972}:

    \begin{equation}
    t=310\left[a+\left(\frac{R}{R_g}\right)^{\frac{3}{2}}\right] M_7\  (s)
    \label{formulabardeen}
    \end{equation} 

Given the BH mass of NGC~3783 of 3$\times$10$^7$ M$_{\odot}$ \citep{peterson2004} and using a spin value $a=0$, we find that a time resolution of 5 ks corresponds to the orbital period at $\mathrm{\sim3\ R_{g}}$. This carries the potential to oversample periodicities or non-random variations occurring at larger radii, meaning that we may probe and map regions just outside of the event horizon. We note that, in terms of light travel time, 5ks correspond to a few tens ($\sim$35) $R_{g}$.
Then we verified that using this duration as a time resolution allows us to have a good photon statistics. To this aim, we extracted spectra in time bins of 5 ks; to give an idea of the available number of counts, the spectrum with the lowest flux (ninth bin of O1), provides $\sim$3500 photons in the 4-10 keV band. By assuming this time resolution, we obtain a total of 88 spectra, 56 for the unobscured dataset and 32 for the obscured one.

\subsection{Baseline model}
\label{baselinemodel}
We first chose a baseline model for each time-resolved spectrum to describe the broad-band continuum. Focusing on the 4-10 keV energy band allows us to apply a simple model, that only includes a power law, an absorption component (intrinsic to the source, as the Galactic absorption does not have effects at these energies), and a narrow Gaussian emission line for the neutral Fe K$\alpha$ line. 
A combination of cold, mildly ionized, and partial covering absorbers has been detected in this source (R04, \citealt{yaqoob2005}, M17, \citealt{mao19}); these absorbers were particularly intense during the obscured observations (M17) \footnote{We note that multilayers of partially-covering obscurers at mild-to-neutral ionization state have been extensively found in Seyfert 1 galaxies in the last decade \citep[e.g.][]{turner09,longinotti2009,reeves2014,kaastra14,cappi16}. These absorbers are in addition to the multicomponent warm absorber which are typically total covering and highly ionized \citep{blustin05,piconcelli05,mckernan07,laha14}.}. However, the duration of our time slices prevents us from testing such complex scenarios. To embrace at least one of the aspects that is in common among all the absorption schemes presented in literature, we thus decided to adopt a partial covering model due to cold matter. This, in principle, should allow the fitting to mimic, if necessary, smoothed curvatures due to either ionized or cold absorbers. On the one hand, far from being a totally realistic representation of what is really happening in the source, our choice is meant to reduce as much as possible any systematics due to an extremely oversimplified modeling of the continuum, at the cost of one single degree of freedom. On the other hand, since our choice intrinsically includes the possibility of having a total obscurer, it is consistent with works where similar techniques were adopted \citep{tombesi2010, gofford2013}. Morever, models of ionized gas were not chosen here because they typically include discrete absorption features, which are those we aim to study separately and individually. As for the iron line, in NCG~3783 this feature is known to be present (R04, \citealt{tombesi2007}, M17, DM20) and is typically sufficiently strong (EW $\sim 100-150$ eV) to alter the 4--10~keV continuum fit if not properly taken into account. 
We define this as our baseline model, i.e. in XSPEC terminology \texttt{pcfabs $\times$ (power law + gauss)}.

 \subsection{Fe K$\alpha$ line}
 \label{fekalphaline}
 Following the results of M17, we assume the Fe K$\alpha$ to be narrow, with a frozen width of 10 eV. The other line parameters (energy and normalization) are left free to vary.
 \begin{figure*}
	\centering
	\resizebox{\hsize}{!}{
	\includegraphics[height=0.25\hsize]{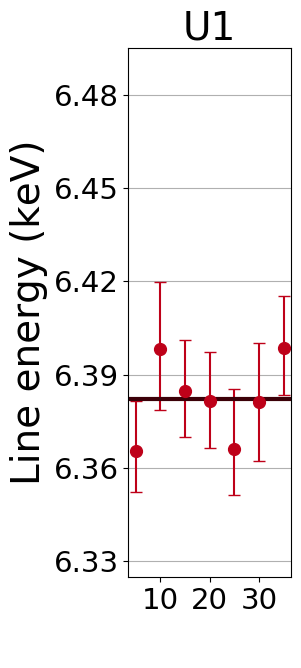}
	\includegraphics[height=0.25\hsize]{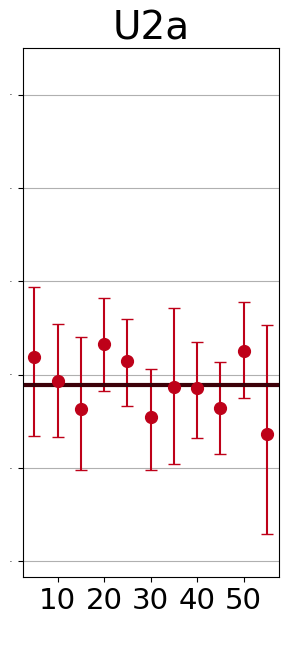}
	\includegraphics[height=0.25\hsize]{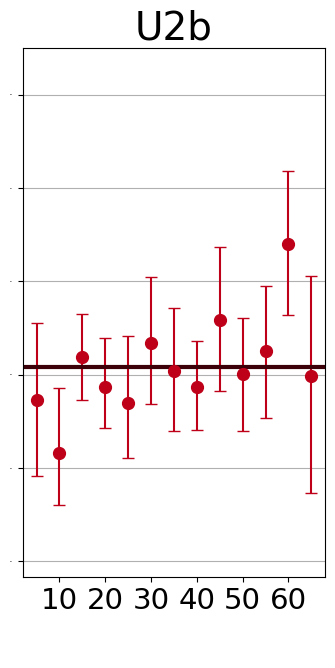}
	\includegraphics[height=0.25\hsize]{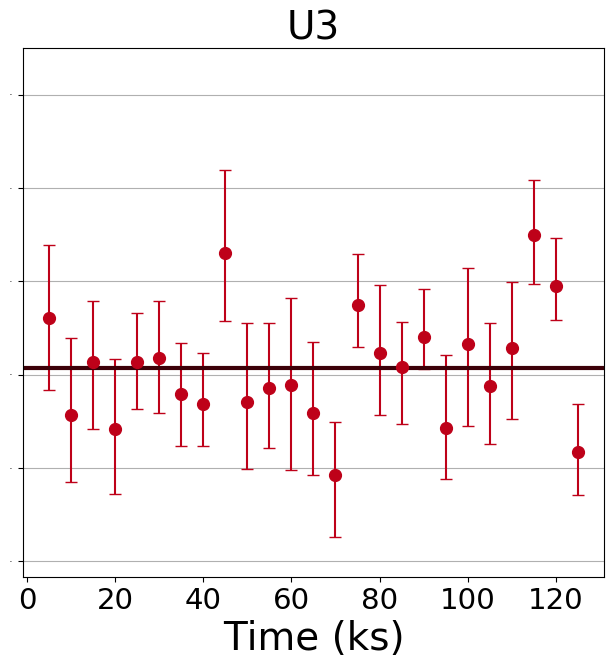} 
	\includegraphics[height=0.25\hsize]{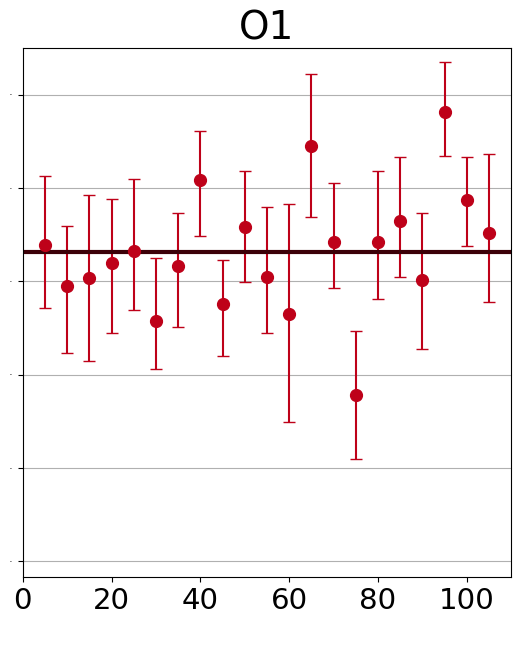}
	\includegraphics[height=0.25\hsize]{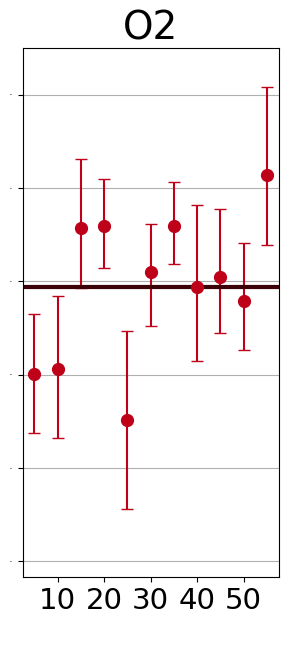}}
	\resizebox{\hsize}{!}{
	\includegraphics[height=0.25\hsize]{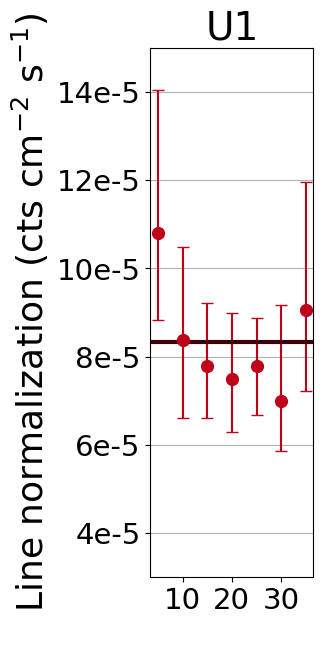}
	\includegraphics[height=0.25\hsize]{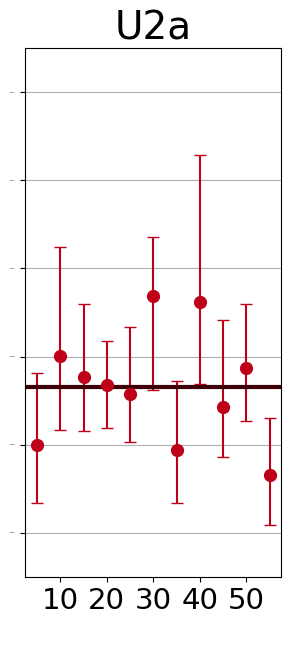}
	\includegraphics[height=0.25\hsize]{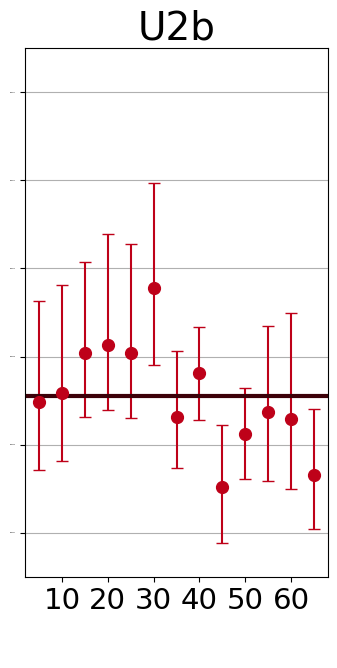}
	\includegraphics[height=0.25\hsize]{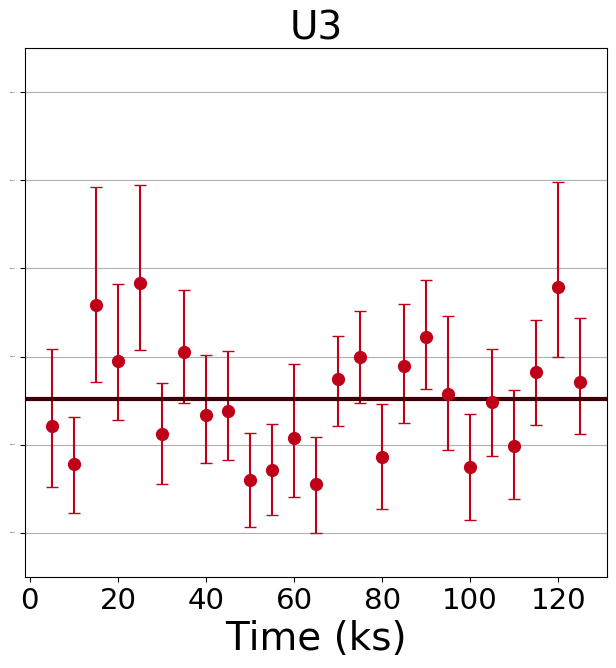}
	\includegraphics[height=0.25\hsize]{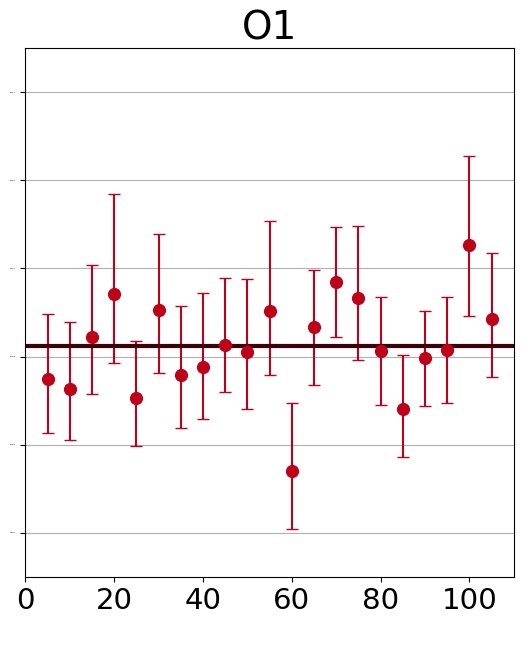}
	\includegraphics[height=0.25\hsize]{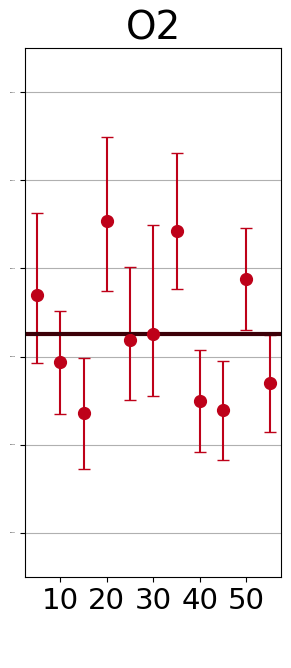}}
    \centering \resizebox{\hsize}{!}{\includegraphics[height=0.25\hsize]{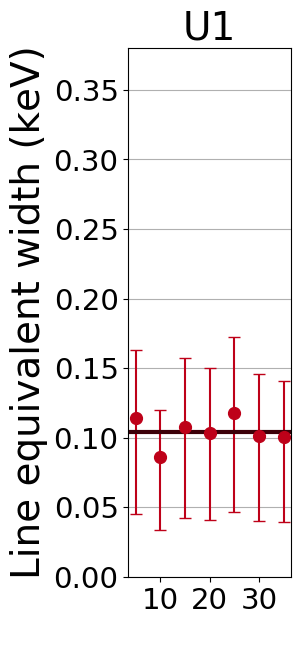}
	\includegraphics[height=0.25\hsize]{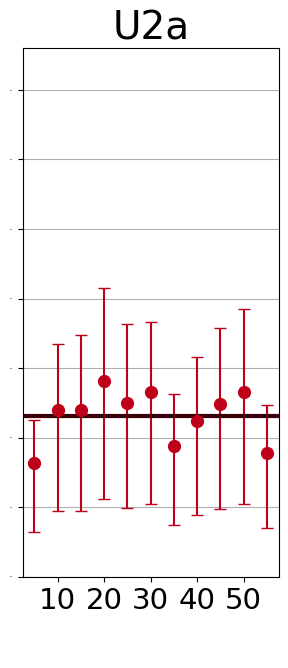}
	\includegraphics[height=0.25\hsize]{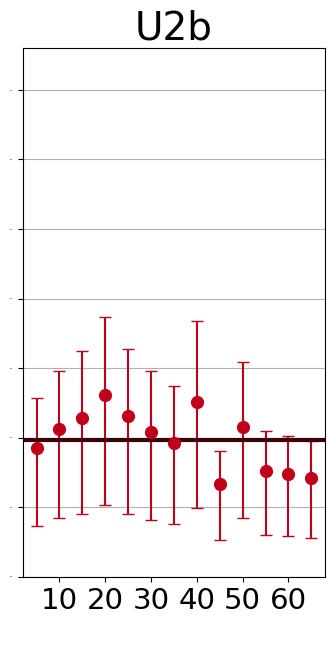}
	\includegraphics[height=0.25\hsize]{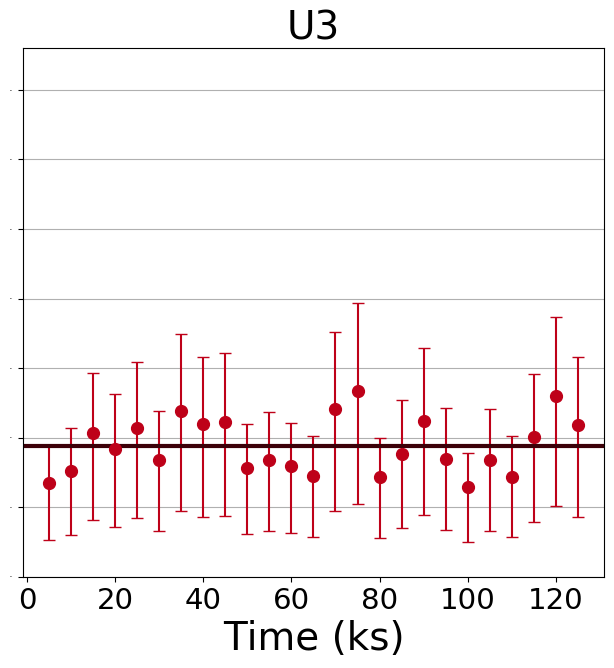} 
	\includegraphics[height=0.25\hsize]{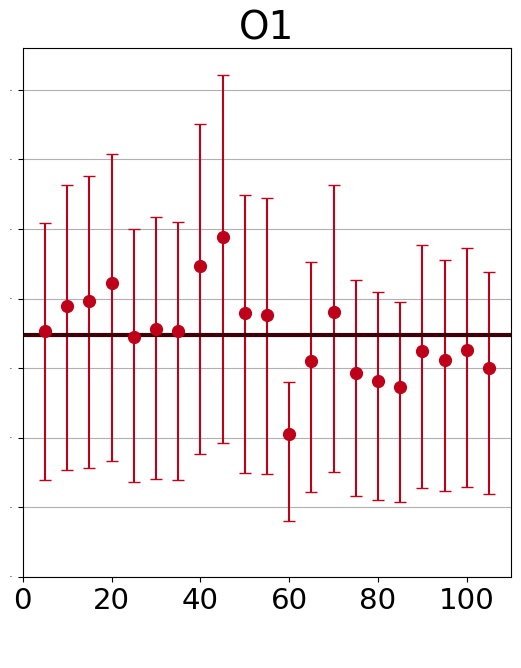}
	\includegraphics[height=0.25\hsize]{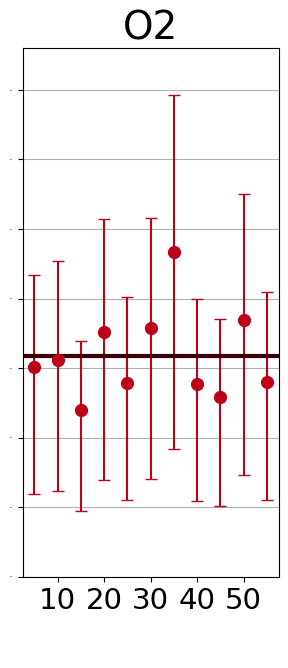}
	}
	\caption{Rest-frame energy (top panel), normalization (middle panel), and EW (bottom panel) of the narrow Fe K$\alpha$ line included in the baseline model; errors are reported at 1$\sigma$.}
	\label{kalfa_energy_eqw}
\end{figure*}

In Fig. \ref{kalfa_energy_eqw} we show the energy, normalization, and equivalent width (EW) of the narrow Fe K$\alpha$ as obtained from the fits of each time-resolved spectrum, and their average values in individual observations. 
We do not observe major variability in the line parameters, apart from a shift in the centroid energy ($\sim$50~eV) between unobscured and obscured data. We checked if a similar behaviour is measurable also in the MOS data using the average spectra for each of the five observations but we did not find any evidence for such a shift. We therefore conclude that the observed shift is most probably due to uncorrected CTI evolution that translates into a poor gain calibration, as described in \cite{ponti13}, \cite{mehdipour2015}, and \cite{zoghbi19}.  
In addition, the energy of the narrow Fe K line displays some short episodes of significant variation (see U3, O1 and O2 in the upper panels of Fig.~\ref{kalfa_energy_eqw}) on time-scales of tens of ks.

Interestingly enough, the line EW seems to be constant within each single pointing, despite the almost $30\%$ variability in continuum flux for the most extreme cases (Fig.~\ref{lightcurves}, U2b, U3, O1). This is not expected if we assume that the narrow Fe K$\alpha$ emission line is produced far from the origin of the primary X-rays. In this case, we would expect a decoupling between the two quantities due to the time-delays introduced by the distance of the reprocessor. The easiest way to explain what is presented in the lower row of Fig.~\ref{kalfa_energy_eqw} is to associate this emission component with a feature produced in the vicinity (i.e. fast responding) of the SMBH. Under this assumption, one would expect it to be related to the relativistically modified iron emission line. On the one hand, of this were the case, we would expect some variability in line shape on short time-scales, as predicted for example in light-bending scenarios \citep{miniutti04}. On the other hand, D20 demonstrated that the variability of the source is strongly influenced by changes in the absorber characteristics also at small time-scales ($\sim1500 s$). This is in agreement with the apparent steadiness of the EW of the emission feature. Overall, we stress here that the 1$\sigma$ error on the Fe K$\alpha$ line normalization is of the order of $25-30\%$ (middle row of Fig.\ref{kalfa_energy_eqw}). This heavily impacts on our capability to deeply investigate the iron line intensity variability on these time-scales. We cannot claim strong evidences for variation but, at the same time, we cannot exclude them if they are of the order of $\sim10\%$.\\On larger time-scales, from 2001 to 2016 observations, we may  appreciate an increase in the average value of Fe K$\alpha$ EW of the order of 70~eV. Assuming that the energy of the line remains the same between the two sets of observation, this increase in EW is consistent with the rise of the absorber column density and decrease in flux (M17). 

\begin{figure*}
	\centering
		\centering
	\resizebox{\hsize}{!}{
		\includegraphics{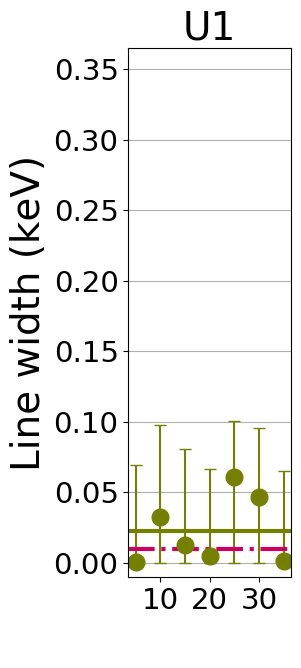}
		\includegraphics{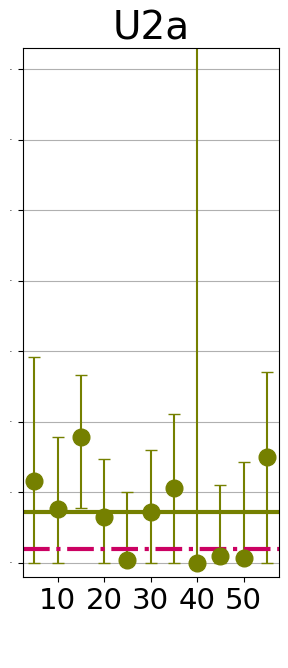}
\includegraphics{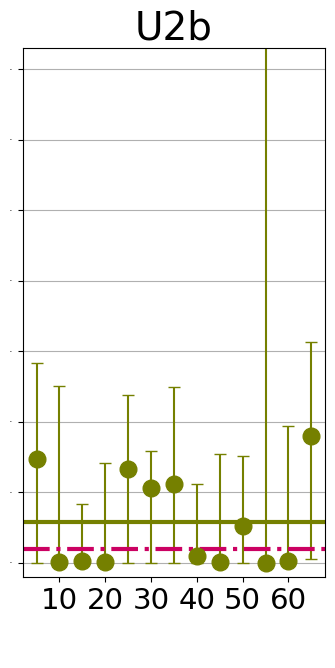}
		\includegraphics{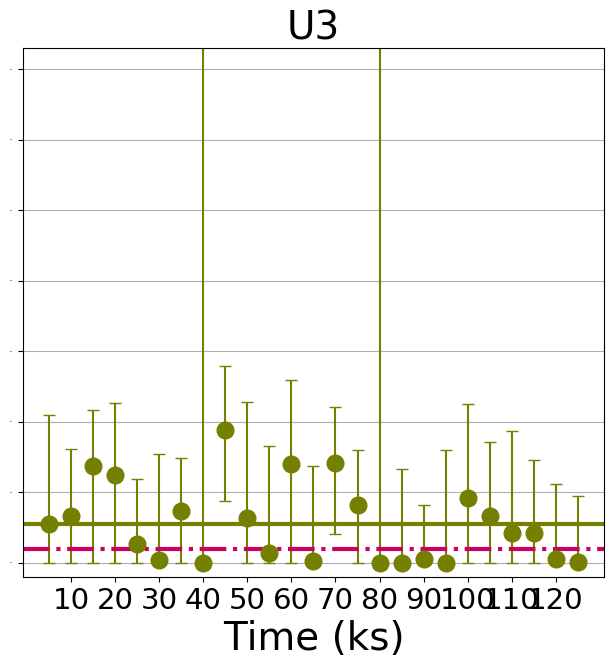} 
		\includegraphics{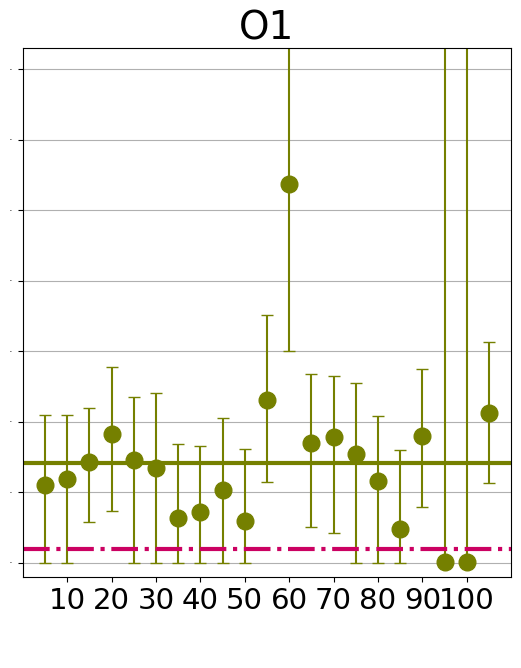}
		\includegraphics{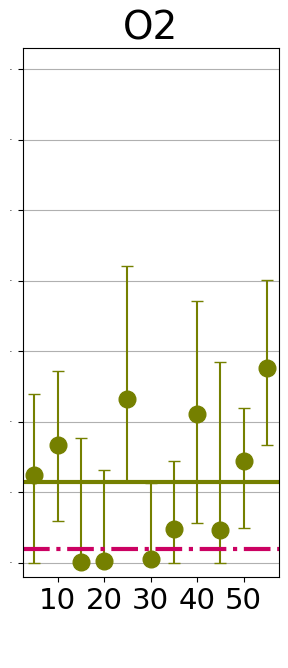}
	}
	\caption{Width of the Fe K$\alpha$ line. The magenta dotted line marks the frozen 10 eV width used in the baseline model. All errors are reported at $1\sigma$.}
	\label{kalfa_en_free_width}
\end{figure*}

We then verified whether the assumption of a narrow line (width fixed at 10 eV) was correct, and let its width free to vary. We obtained  average values of its width of $\sim$30 eV for the unobscured state and $\sim$70 eV for the obscured state (Fig.~\ref{kalfa_en_free_width}). The width was, however, always consistent with 0 to within 2 $\sigma$. We thus decided to keep the width frozen to 10 eV for the following steps of the analysis. 

\subsection{Blind search}
\label{blindsearch}

\begin{figure}
		\centering
		\includegraphics[width=\hsize]{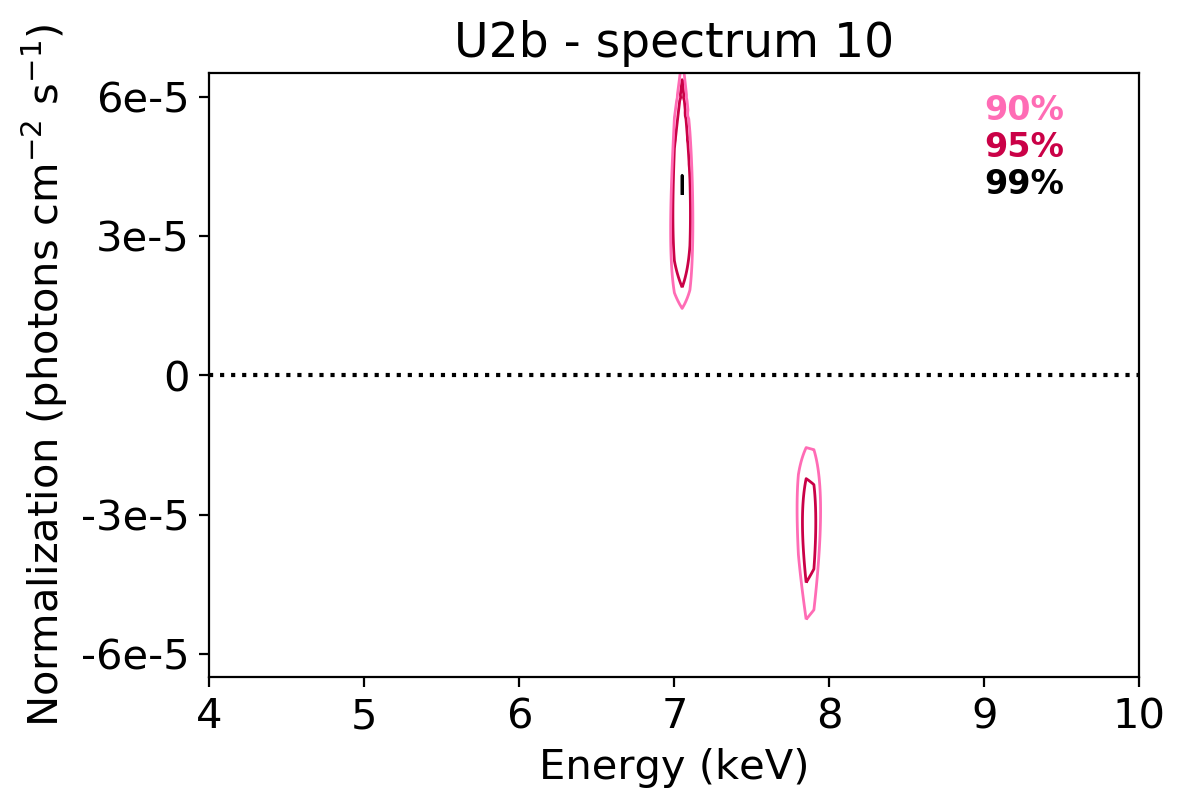}
		\caption{An example of the significance contours found with the blind search in a single 5~ks spectrum. An emission line is detected at $\mathrm{\sim3\sigma\ (99\%,\ i.e.\ \Delta\chi^2\leq-11.34,\ \Delta d.o.f.=3)}$ at $\sim6.9$ keV and an absorption line is detected at $\mathrm{\sim2\sigma\ (95\%,\ i.e.\ -7.81\geq\Delta\chi^2> -11.34,\ \Delta d.o.f.=3)}$ at $\sim7.9$ keV.}
		\label{esempiocontorni}
	\end{figure}

After fitting the baseline model described in Sect.~\ref{baselinemodel}, we carry out a blind search for additional emission/absorption features. Since the only discrete component present in the baseline model is the Fe K$\alpha$, we may expect to detect also features (like the Fe K$\beta$/ionized Fe K$\alpha$ blend and the absorption lines described in R04 and M17) that appear in the average spectrum. Our purpose is to verify whether they are present at all times or do show some variations.
We apply a procedure similar to that adopted in \cite{tombesi2010}: a second Gaussian component is included in the model, allowing for both positive and negative values of the normalization, and with width in the range 0.01--0.5~keV. Then the \texttt{steppar} command is launched simultaneously on the line energy parameter (which can vary from 4 to 10~keV, with increments of 5 eV) and the normalization parameter (from -6.5 to +6.5 $\times\ 10^{-5}$ photons/s/cm$^{-2}$, with increments of $6.5\times\ 10^{-7}$ photons/s/cm$^{-2}$). We then plot the significance contours corresponding to $\Delta\chi^2$ of -6.25, -7.81, -11.34 that, for three free parameters, represent a significance of the line of 1.6 $(\geq90\%)$, 2 $(\geq95\%)$, and 3 $(\geq99\%)$  $\sigma$, respectively. An example of these contour plots is shown in Fig.~\ref{esempiocontorni}, where an emission line is detected at $\sim3\sigma$ at E$\sim$6.9~keV and an absorption line is found at $\sim2\sigma$ at $\sim$7.9~keV. 
This procedure is repeated for all the 5ks-long, time-resolved spectra.
Since the obscured/unobscured state of the X-ray source may influence the number and type (i.e. emission vs absorption) of lines, we analyzed separately the features detected in the two states.
\begin{figure*}
		\centering
		{\includegraphics[width=.47\textwidth]{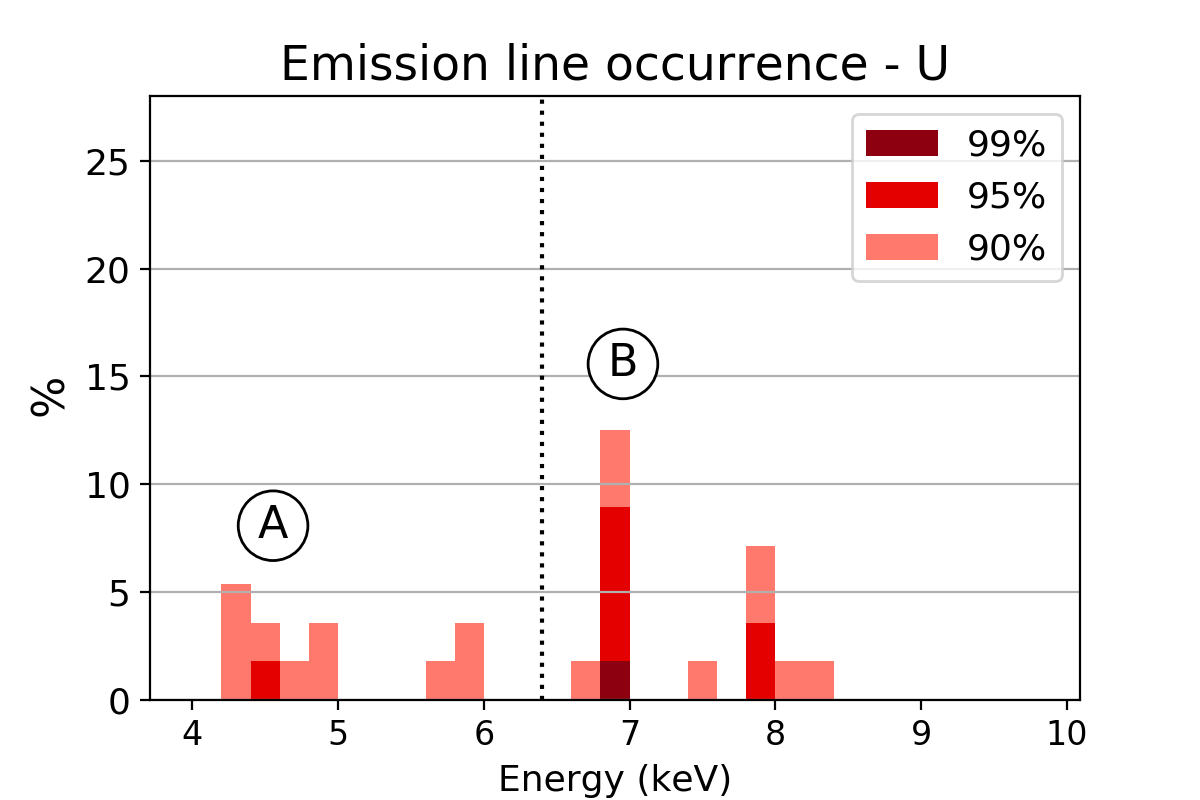}}\quad
		{\includegraphics[width=.47\textwidth]{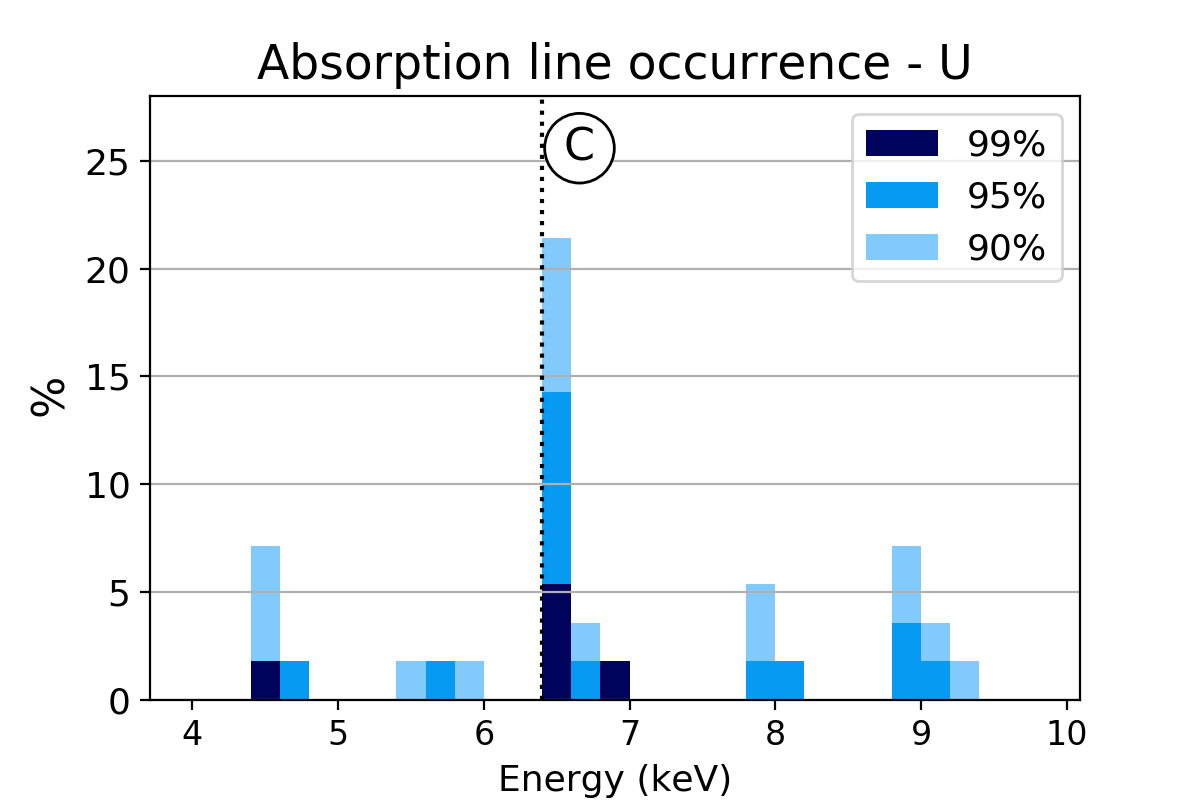}} \\
		{\includegraphics[width=.47\textwidth]{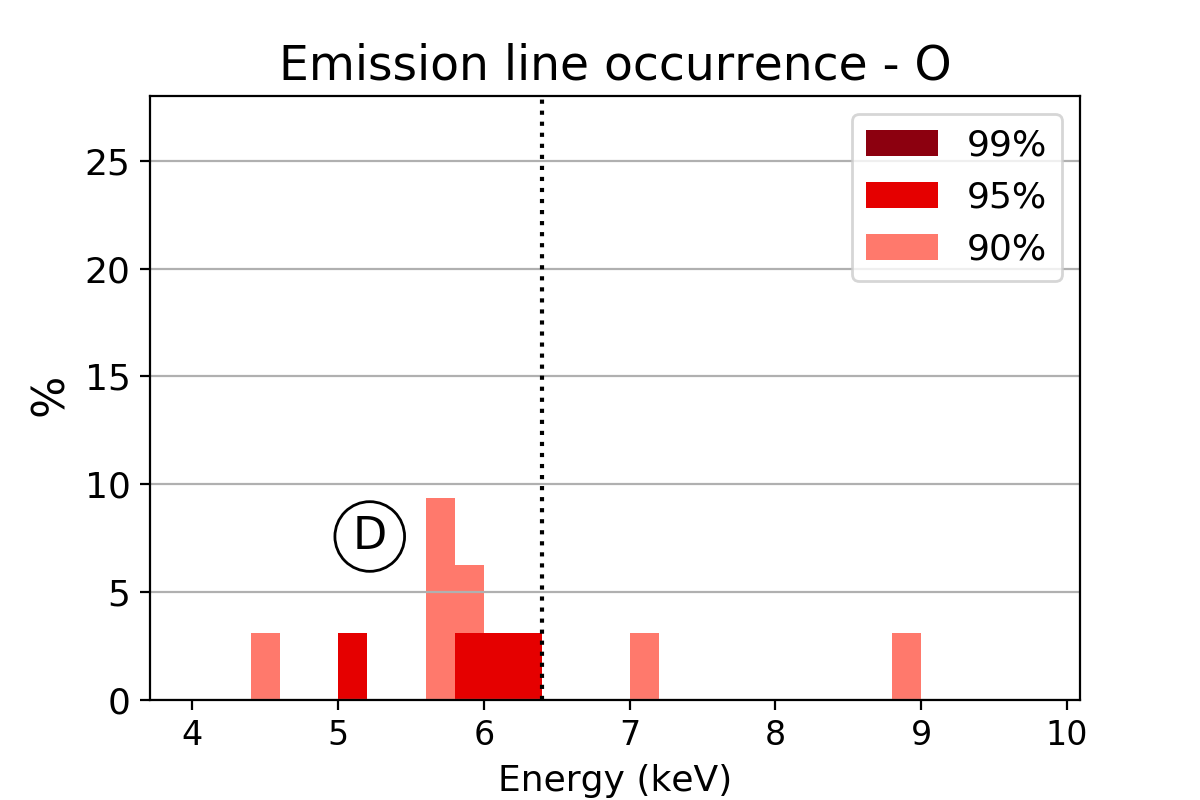}}\quad
		{\includegraphics[width=.47\textwidth]{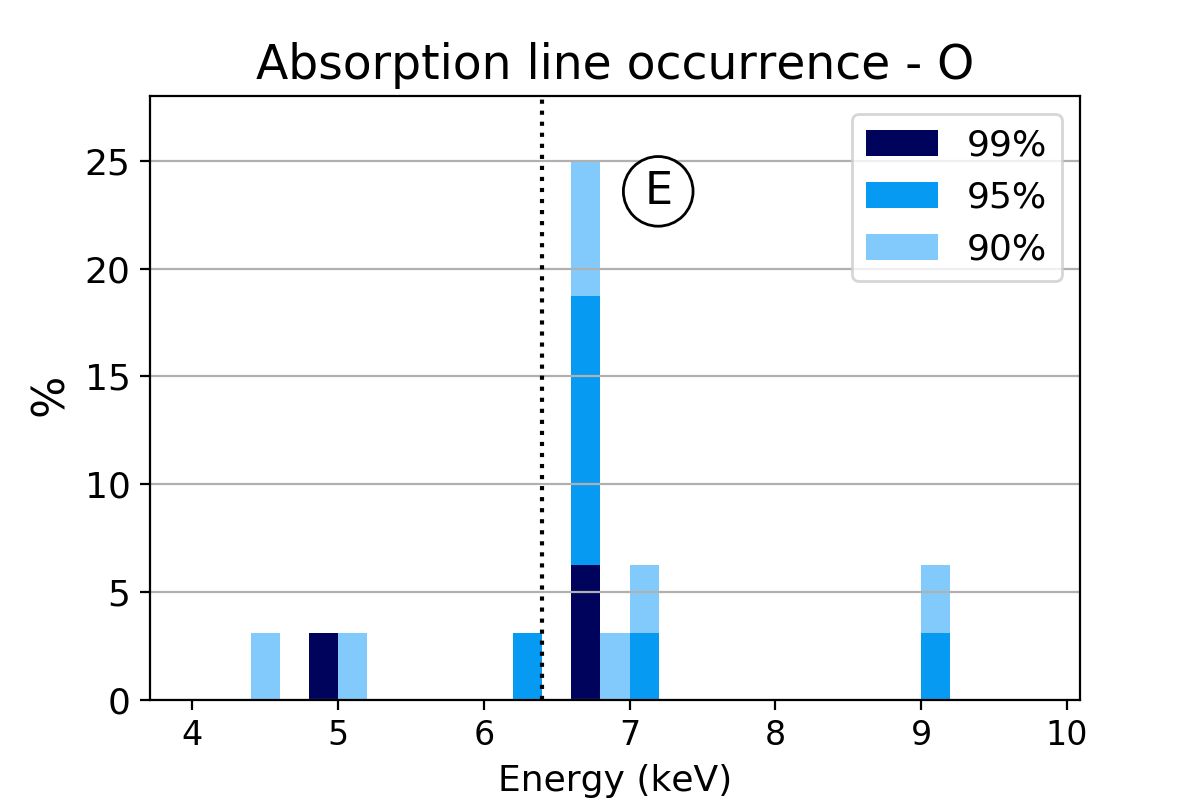}}		
    \caption{The distribution in energy (observed frame) of the features detected via blind search. The energy bins are 0.2 keV wide. The top panels show the results of the unobscured dataset (U1, U2a, U2b, U3), while the bottom panels refer to the obscured datasets (O1, O2). On the left panels, detections of emission lines are reported (red), while the right panels report detections of absorption lines (blue): in both cases, the darker the color the higher the significance, i.e. the measured $\Delta\chi^2$. The percentage on the y axes is the ratio between the number of detections in each bin and the total number of spectra for the two sets of observations (56 in total for the unobscured, and 32 in total for the obscured). The dotted line reported in all four figures is the nominal energy of the Fe K$\alpha$ and is shown purely as a visual reference. The letters A-E indicate the clusters of lines that globally reach a significance $\gtrsim3\sigma$, as indicated in Table \ref{tab_sign} and described in Sect.~\ref{simulations}.   }
		\label{istogrammi}
\end{figure*}
 
In Fig.~\ref{istogrammi} the distributions of the detected emission and absorption features as a function of their observed-frame energy are reported; each bin is 0.2~keV wide. The vertical dotted line represents the nominal energy of the Fe K$\alpha$ emission and is shown only as a visual reference, since in the current analysis this feature is already included in the model as a narrow emission line (see Sect.~\ref{fekalphaline}). The occurrences of the features are normalized to the number of analyzed spectra, 56 for the unobscured case and 32 for the obscured case.  
Interestingly, the energy of the detected emission and absorption lines seems to cluster around certain values. 
Starting with the emission lines, in the unobscured data we recognize a small peak in the distribution (corresponding to a total of 8 detections in time-resolved spectra) at the energy of the K$\beta$/ionized K$\alpha$ line, corresponding to the K$\beta$/ionized K$\alpha$ blend. In the obscured state (Fig.~\ref{istogrammi}, bottom-left), most of the detected lines are clustered between $\sim5.6$ keV and $\sim$6.4 keV.\\
For the absorption features, most of the detections cluster between $\sim$6.4 keV and $\sim$7 keV. While no major difference is observed in the distribution of clustered features in unobscured and obscured observations, there seems to be a slight shift towards higher energies in the obscured datasets.

\subsection{Simulations and significance of the features}
\label{simulations}
To assess the statistical significance of the individual lines, we followed the indications reported by \cite{protassov2002} and \cite{vaughanuttley2008}. In particular, after a preliminary assessment by evaluating the $\Delta\chi^2$ for the 3 $\Delta$d.o.f. of the detected line (energy, normalization, width), we used Monte Carlo techniques to compute the number of times a line at a given energy and significance is found by chance in simulated, featureless spectraIn doing so, we took advantage of what we found in real data: in fact, the detected features we are looking for in the simulations do not appear to be random fluctuations, that could be positive or negative, but lines clustered in determined regions of the spectrum, bound to be strictly positive or negative. We thus tested separately emission and absorption lines. \\
To probe the different source states, we analyzed separately the unobscured and obscured epochs, because of their intrinsic diversity, and for each epoch we made a selection in X-ray flux, identifying low, medium and high-flux states. For the 2000/2001 datasets, the flux ranges are 2.50$-$3.09, 3.09$-$3.68, 3.68$-$4.27 $\times \  10^{-11}$ erg s$^{-1}$ cm$^{-2}$, while for the 2016 dataset the three flux intervals are 1.42$-$1.80, 1.80$-$2.19, 2.19$-$2.57$\times \ 10^{-11}$ erg s$^{-1}$ cm$^{-2}$, respectively. 
We determined a best-fit baseline model (using the same continuum model defined in Sect.~\ref{baselinemodel}) for each of these groups of spectra (six in total), and used it to simulate 1000 spectra for each group. 
Simulated spectra are grouped at 1/5 of the instrument energy resolution. We first apply and fit the baseline model and record the obtained best fit $\chi^2$. Then a second Gaussian component is added. Since in real data the detected lines appear to cluster at specific energies, we analyzed the energy bands 4-6 keV, 6-8 keV, and 8-10 keV separately in simulated spectra. In particular, the fitted line energy is forced to vary within each of these intervals, and its width is free to vary in the range 0.01--0.5~keV. Emission and absorption features are searched separately.
To adopt a conservative approach, for each cluster of features we finally counted how many of the 1000 trials show a $\Delta\chi^2$ higher than the lowest value found in the actual data. This does automatically translate into a robust estimate of the significance of each of the detected lines. Results are shown in Table~\ref{tab_sign}.

We find that both emission and absorption lines do not reach a high significance in the unobscured dataset and that they tend overall to be more significant during the obscured phase. We also find that the flux state does not seem to have a direct effect on the significance of the features.
If we take into account their occurrences and calculate the binomial probability, their significance increases sensibly, as shown in Tab.~\ref{tab_sign}, where the most significant groups of features are identified with the A-E letters in the last column. The global probability that the absorption lines detected in the 6-8~keV range are not fluctuations is $\gtrsim 4\sigma $ for both source states (C, E).  As for the emission lines, their estimated significance is $\gtrsim3\sigma$ in the 4-6 keV range in all the observations (A, D), and 4$\sigma$ in the 6-8 keV band during the unobscured state (B). 
 
The transient/variable nature of the detected features is naturally inferred from the fact that they are detected only in a fraction of our time-resolved spectra.

\begin{table}
	\centering
	\begin{tabular}{c|c|ccc|c|c}		
	\toprule
	Energy& $\Delta\chi^2\ \sigma_{single}$ \tiny{(1)} & \multicolumn{3}{c|}{MC $\sigma_{single}$ \tiny{(2)}} & $\sigma_{group}$ \tiny{(3)}&\\
	&&\tiny{Low} & \tiny{Mid} & \tiny{High} &&\\
	\tiny{keV} &  \tiny{\%}&\tiny{\%} &\tiny{\%}&\tiny{\%}  & \tiny{\%}\\
		\midrule
		\multicolumn{7}{c}{Unobscured} \\ 
		\hline
		\multicolumn{7}{c}{Emission} \\
		\hline
	    4-6&91&94&92&92&99.3&A\\
    	6-8&91&94&93&93&>99.9&B\\
	    8-10&91&92&90&91&91.0&\\
    	\midrule
		\multicolumn{6}{c}{Absorption} \\
		\hline
		4-6&92&90&91&88&87.0&\\
		6-8&91&90&88&89&99.9&C\\
		8-10&90&87&83&80&91.0&\\
		\midrule
		\multicolumn{7}{c}{Obscured} \\ 
		\hline
		\multicolumn{7}{c}{Emission} \\
		\hline
		4-6&95&97&96&96&>99.9&D\\
		6-8&96&99&99&98&99.0&\\
		8-10&94&96&96&95&65.0&\\
		\hline
		\multicolumn{7}{c}{Absorption} \\
		\hline
	    4-6&97&97&96&97&93.0&\\
		6-8&92&91&93&92&>99.9&E\\
		8-10&96&96&95&95&74.0&\\
		\bottomrule
	\end{tabular} 
	\caption{Significance of detected variable features: (1)\ significance of single features calculated from the $\Delta\chi^{2}$ for 3 $\Delta$d.o.f. measured after the addition of a Gaussian line in emission/absorption; (2) significance of single features from MonteCarlo simulations (Low/Mid/High refer to the different selected flux levels, defined in Sect.~\ref{simulations}); (3) global significance of detected features (i.e. calculated from a binomial distribution, and assuming the total number of spectra in each state as the number of trials, the number of detections obtained via the blind search as the number of successes, and the mean value from MC simulations as the success probability for each trial). The letters A-E refer to the groups of features with a significance $>3\sigma$, as indicated in Sect~\ref{simulations} and displayed in Fig.~\ref{istogrammi}. Their nature is discussed in Sect.~\ref{discussion}.}
	\label{tab_sign}
\end{table}
\section{Residual maps (RM)}
\label{residualmaps}
The blind search can accurately detect features in each time bin and provides information about their possible repeated appearances at different times during the whole observations. It is not simple, though, to understand the evolution in time of such features just from the blind search results: an easy way to trace it is to represent these features in a time-energy plane.
This approach was first introduced by \citet{iwasawa2004} and later used in several studies (e.g. \citealt{turner2006,tombesi2007,demarco2009,nardini2016,marinucci2020}).
Since we have studied the presence of emission and absorption features simultaneously, we decided to visualize them altogether (hence the passage from \textit{excess} to \textit{residual} maps in the denomination).

In order to visualize the data uniformly, we have to choose first an energy resolution for the maps. To have sufficient statistics, we impose a minimum of 20 photons in each $\mathrm{\Delta t\cdot\Delta E}$ pixel. Having already selected $\Delta$t=5~ks (see Sect.~\ref{variablefeaturessearch}), we can adopt $\Delta$E = 100~eV, which is approximately equal to the EPIC pn energy resolution at high energies and allows us to to collect >20 photons per time-energy pixel at all times, except for energies higher than 9~keV in the first half of O1 (where the flux is at its minimum, as shown in Fig.~\ref{lightcurves}). However, with the blind search we already found that we are not able to detect significant features in this range for the obscured dataset, therefore we decided to limit our analysis to the data below 9 keV for the RM analysis. 

\begin{figure*}
	\centering
	\resizebox{\hsize}{!}{
		\includegraphics{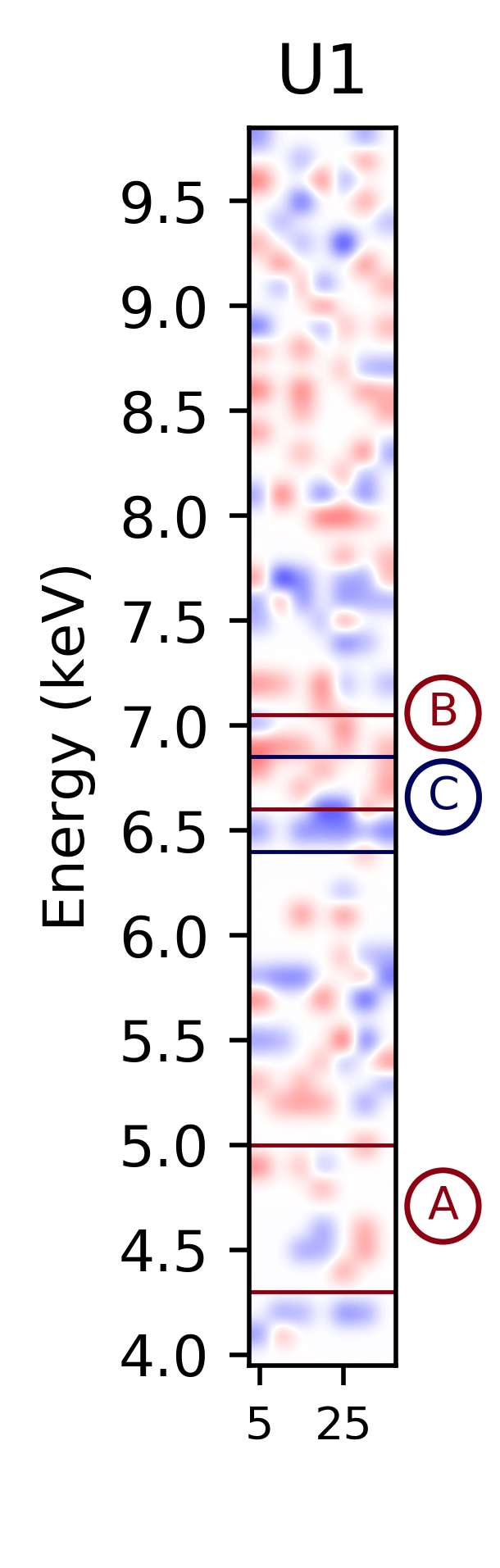}
		\includegraphics{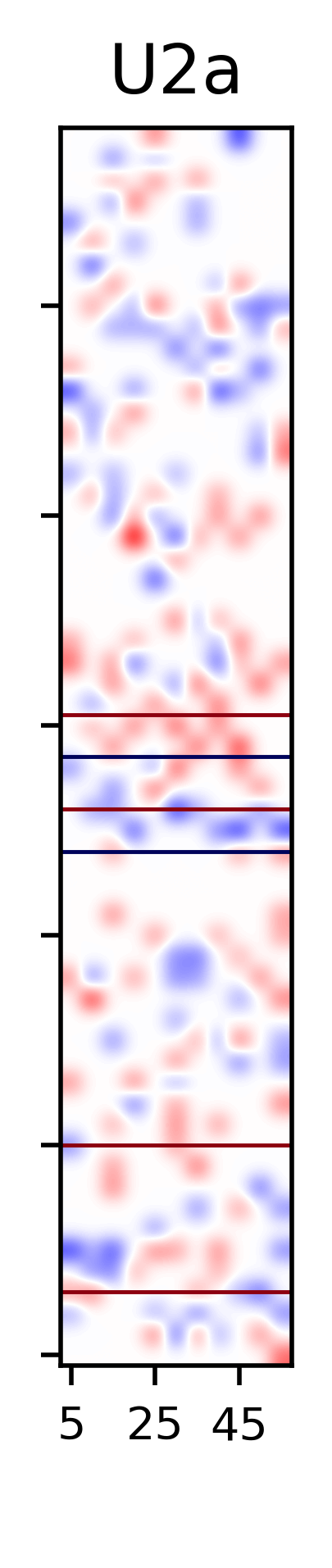}
		\includegraphics{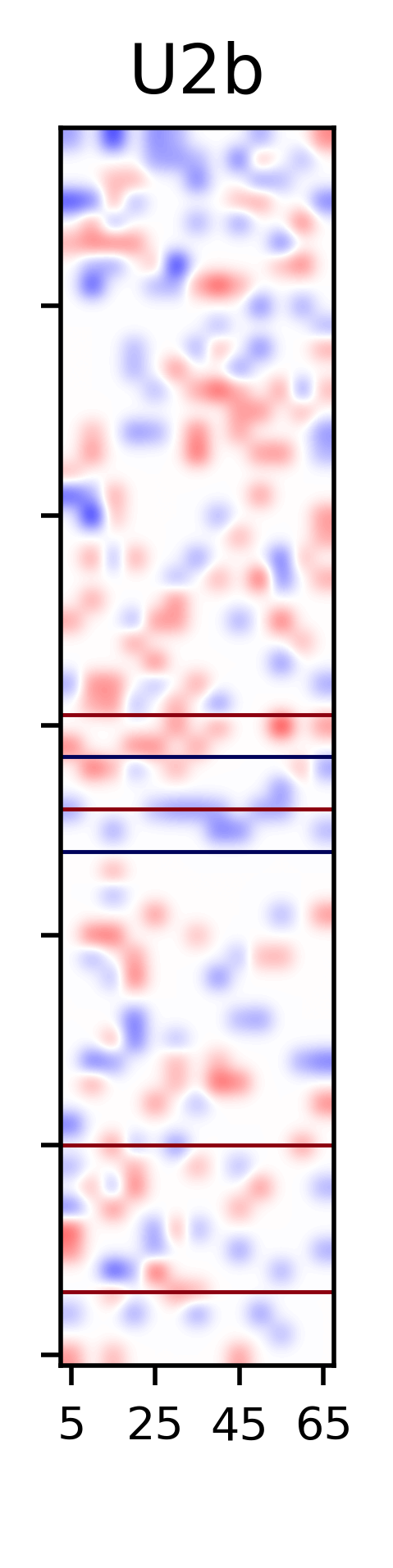}
		\includegraphics{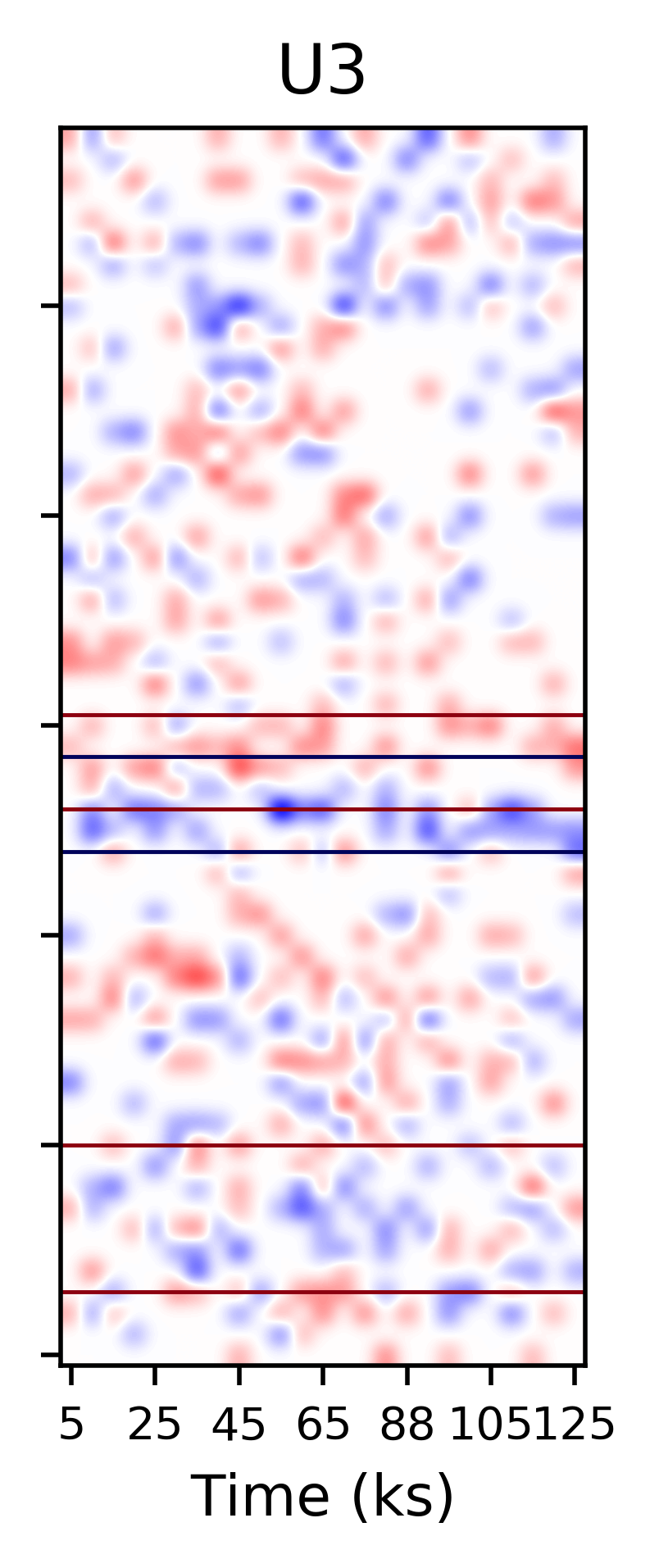}
		\includegraphics{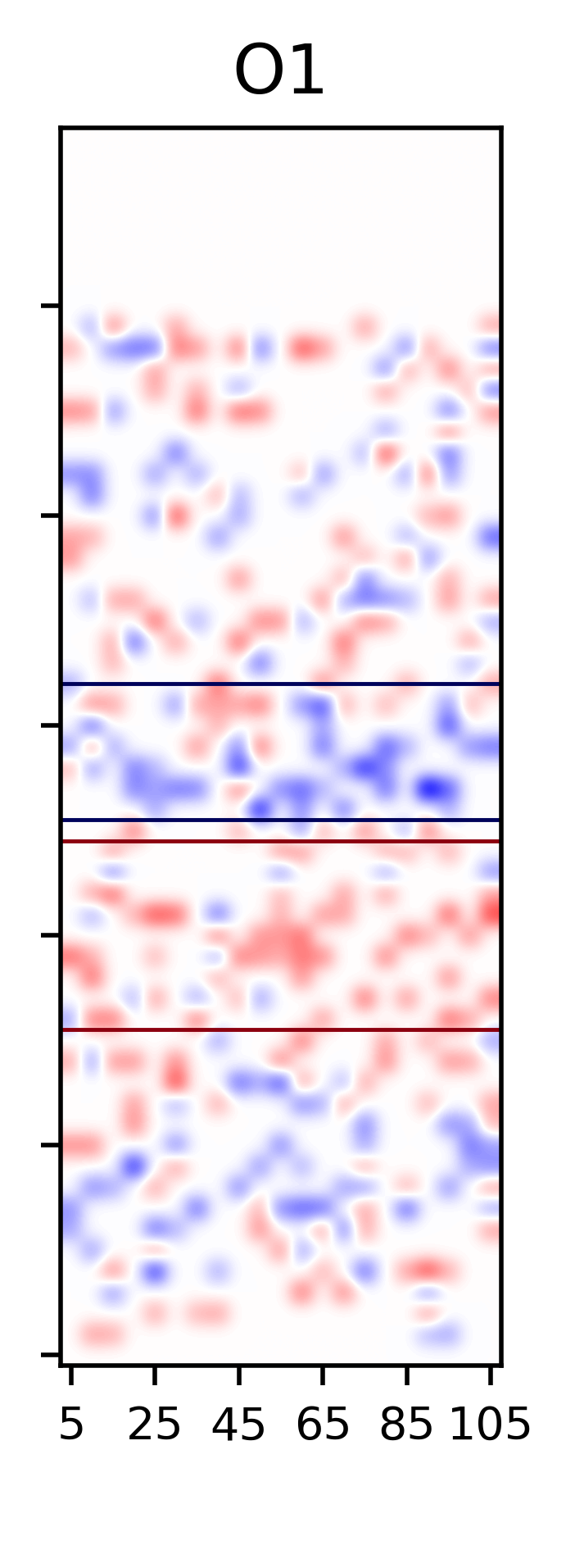}
		\includegraphics{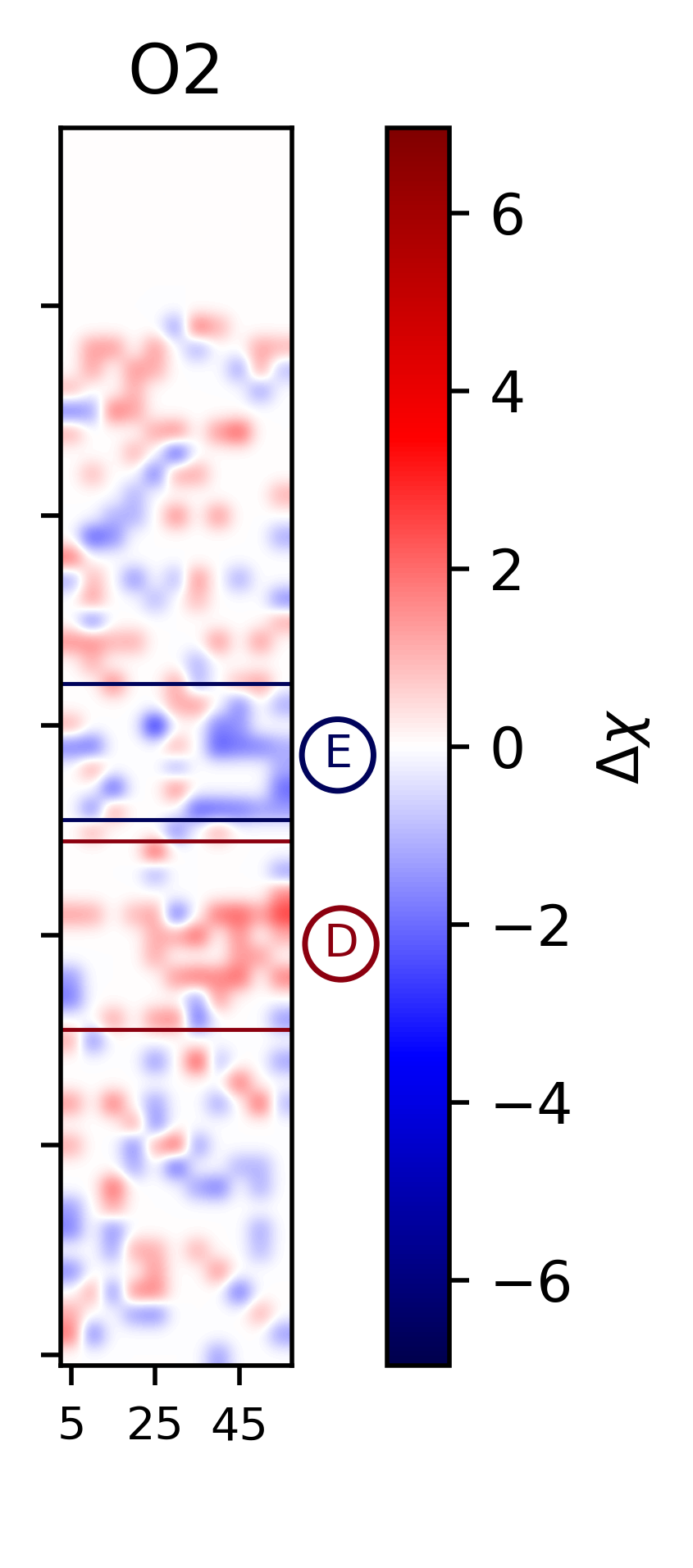}
		
	}
	\caption{Residual maps produced with the  baseline model \texttt{pcfabs $\times$ (power law + gauss)}. The time bin size is 5~ks, while the energy resolution is 0.1~keV. Each pixel shows the value of $\Delta\chi$ (i.e. residual/data error, in red for the positive and in blue for the negative residuals) to give an estimate of the significance, even though the RM are meant to be used here as a qualitative way to identify patterns, rather than a tool to estimate the significance of each feature. A Gaussian interpolation is applied to the pixels to highlight the features and facilitate the identification of possible patterns. The area at E > 9 keV of both O1 and O2 are purposely left blank as the threshold of 20 photons per pixel is not met because of the low flux (see Sect.~\ref{residualmaps}).The horizontal lines highlight the energy bands of most significant groups of features, indicated in Fig.\ref{istogrammi} and in Tab.\ref{tab_sign}.}
	\label{pcfabsmaps}
\end{figure*}

Fig.~\ref{pcfabsmaps} shows the RM produced applying the baseline model described in Sect.~\ref{baselinemodel} fitted in the total energy range. This procedure is fairly different from what is usually done in excess maps \citep[e.g.][]{iwasawa2004,demarco2009}, where the continuum is typically modeled using only narrow energy bands of the spectrum where no major discrete features are expected to be observed (see Appendix \ref{appendice}). As for the blind search, however, we found that the oversimplification of the modeling/fitting may introduce too strong systematics also in the production of the maps (see, for example, the differences between the maps shown in Fig. \ref{pcfabsmaps} and in Fig. \ref{appendixmaps}).

From a visual inspection of the RM, it is possible to see that the features detected in Sect.\ref{variablefeaturessearch} (and listed with A-E letters in Table~\ref{tab_sign}) are present also in the maps, and evidently exhibit intensity variability on short time-scales probed by the RM. Nevertheless, we do not find any clear/obvious macro pattern of variability.
%

The most evident features in the RM are, of course, those that are most conspicuous in the histograms in Fig.~\ref{istogrammi}. 
The recurrent absorption feature detected via the blind search is clearly recognizable as a blue stripe at $\sim$6.5 keV in the unobscured datasets (C), with a varying intensity on short ($\sim10$ ks) time-scales. This absorption line gets wider and shifted in energy up to $\sim$6.7-6.9 keV in the obscured dataset (E). It is interesting to see that, in the RM, this feature seems to appear far more often than in the histograms, where it reaches a maximum frequency of $\sim25\%$: this is because in the blind search we set the detection threshold at 90$\%$, whereas the maps show all the residuals with $|\Delta\chi|>1$, that correspond to a lower significance, approximately 68$\%$.

The emission features related to the neutral Fe K$\beta$/ionized Fe K$\alpha$ are visible in the 2000/2001 observations as an irregular sequence of shallow red spots around $\sim7$ keV (B).
The recurrent emission features in the lower energy band (4-6 keV) also appear shallow and irregularly distributed in the unobscured dataset (A), while in the obscured one they are mostly clustered above 5 keV (D), as especially clear in O1.
These results will be discussed in Sect.~\ref{discussion}.

\subsection{Fe K$\alpha$ line residuals}
\label{fekalpharesiduals}
\begin{figure*}
	\centering
	\resizebox{\hsize}{!}{
		\includegraphics{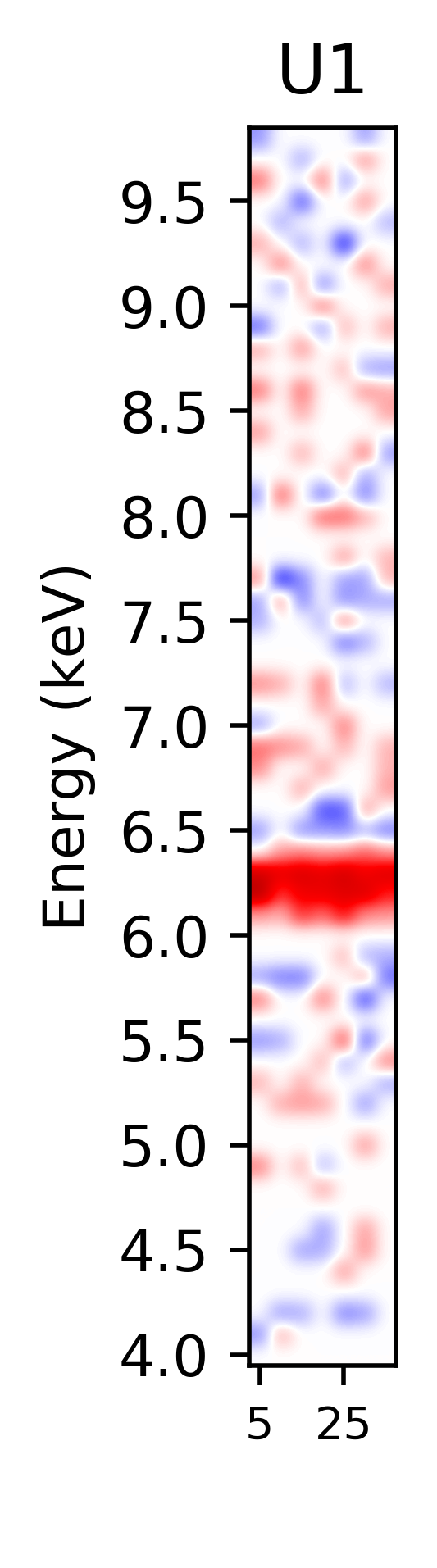}
		\includegraphics{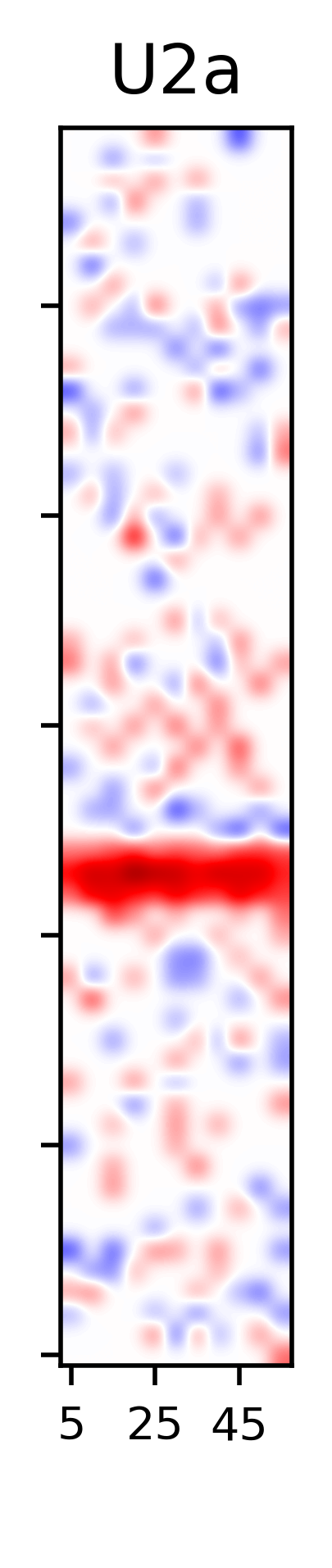}
		\includegraphics{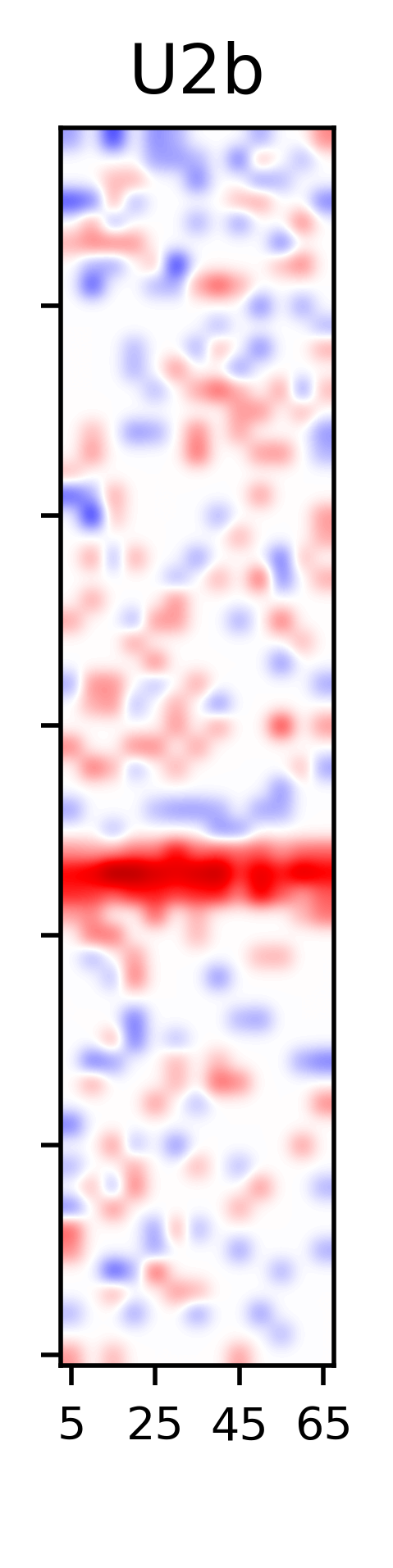}
		\includegraphics{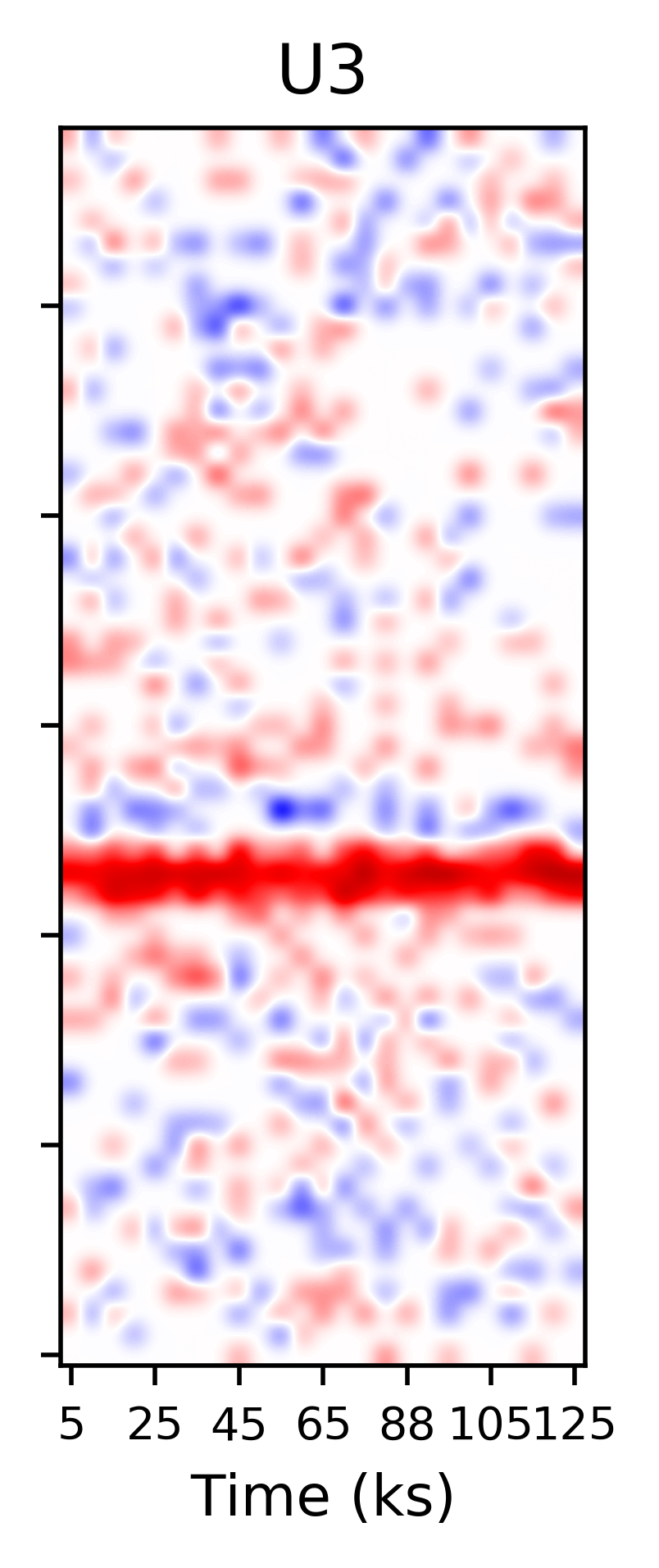}
		\includegraphics{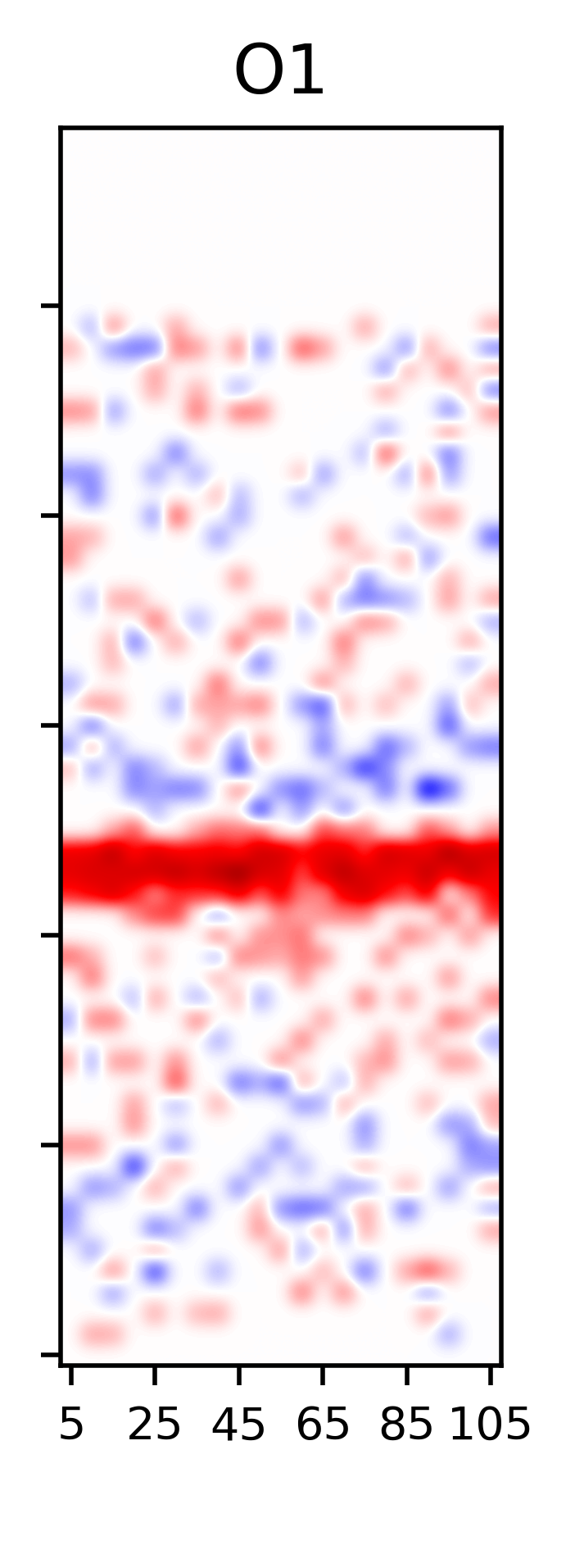}
		\includegraphics{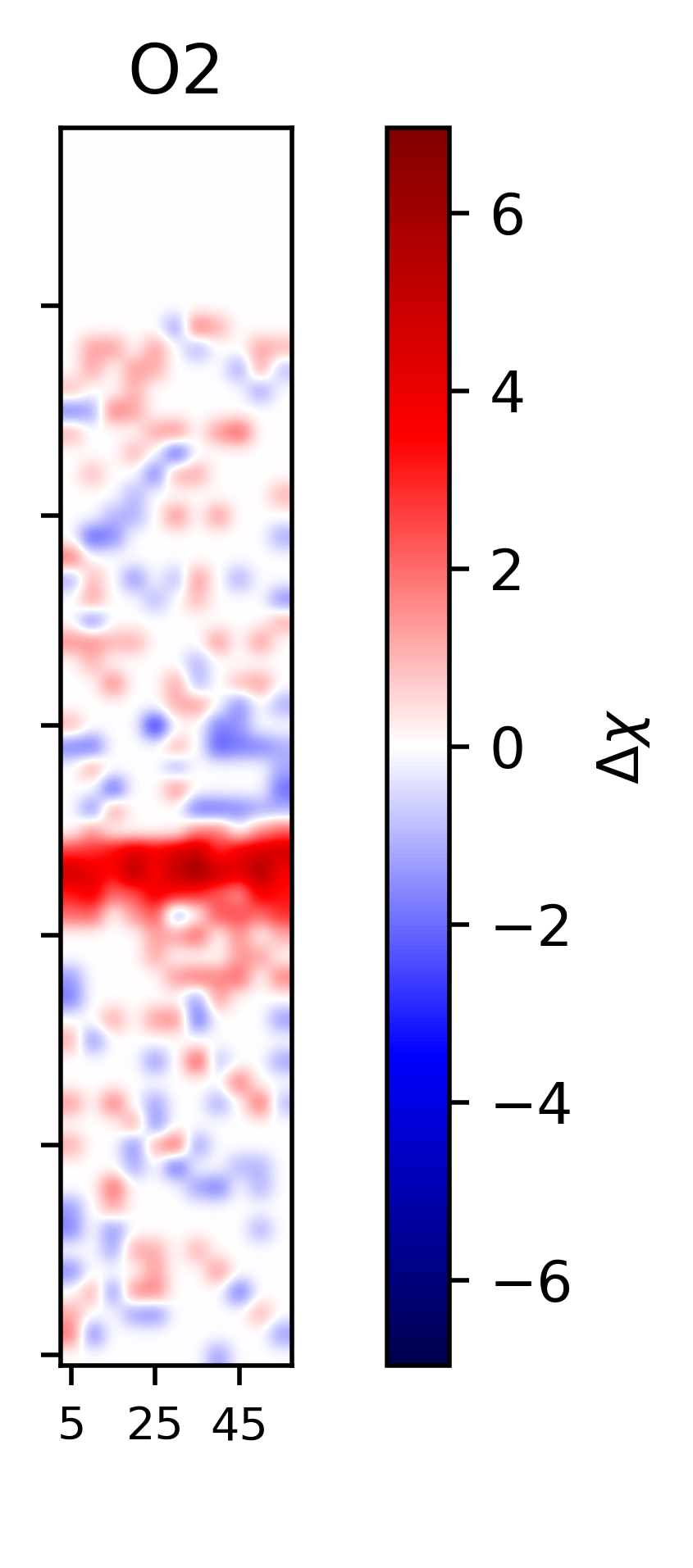}
		
	}
	\caption{Residual maps including the narrow Fe K$\alpha$ component, produced as described in Sect.~\ref{fekalpharesiduals}. These residuals are produced by fixing the baseline model at its best fit values for each spectrum and then removing the Gaussian component.} 
	\label{pcfabslinelinemaps}
\end{figure*}

The RM shown in Fig.~\ref{pcfabsmaps} provide information about all possible lines in the 4-10 keV band except for the narrow component of the Fe K$\alpha$, as it is fitted in each spectrum, so the residuals are null by default. Even though its variations in energy, EW, and width are already shown in Fig.\ref{kalfa_energy_eqw} and Fig. \ref{kalfa_en_free_width}, it would still be interesting to see them along with the other features, to highlight possible relations. Hence we produced a new set of RM, where we fitted the baseline model described in Sect.~\ref{baselinemodel}, froze all the parameters of the absorber and of the power law at their best values, then removed the line component, and finally plotted the $\Delta\chi$. These RM are shown in Fig.~\ref{pcfabslinelinemaps}. The strong emission feature at $\sim$6.4~keV is always present, but the darker and lighter spots seem to indicate some variability. Given the results obtained from the fits of our baseline model (Sect.~\ref{baselinemodel}), the observed variations in the RM can be ascribed to a combination of the variations of EW and width of the Fe K$\alpha$ emission line (Fig.~\ref{kalfa_energy_eqw}, bottom panel; Fig.~\ref{kalfa_en_free_width}). 
The possible origin of this behaviour will be discussed in Sect.~\ref{modulation}. 

\section{Discussion}
\label{discussion}

In this work we searched for and identified absorption and emission features between 4-10 keV that are variable on short time scales in the Seyfert galaxy NGC 3783. We studied archival XMM-Newton observations which caught the source in two different states: one unobscured and one obscured state (as previously found by M17). This allowed us to investigate how these features change between the two states.\\

\textbf{Features in emission:} As displayed in the top left panel of Fig. \ref{istogrammi} (B), about 15$\%$ of the spectra during the unobscured epochs (2000/2001) show the presence of emission features with a significance $\geq90\%$ around 7 keV.
This corresponds to a global statistical significance of the detections of $\sim4\sigma$ (Table~\ref{tab_sign}). Conversely, during the obscured epochs the number of detections and their significance drops: we observe only one feature, detected at a significance of $<3\sigma$.
This cluster of emission lines in the unobscured epoch can be identified as either the neutral Fe K$\beta$ line or an additional ionized Fe component found at $\sim$6.9-7 keV. This identification is supported by the ratio between the residuals at $\sim$7 keV and those at $\sim$6.4 keV, that is about a factor 3 larger than what we would expect in the case of a pure Fe K$\beta$ ($\sim1/9$) \citep{molendi03}. The additional contribution to the emission features at $\sim$7 keV may come from the same ionized medium producing the absorption features. The same conclusion is given in R04. This excess is far less visible, if not absent, in the obscured dataset, as we would expect from the results of M17, who suggested that a juxtaposition with a Fe \textsc{XXVI} Ly$\alpha$ absorption line cancels out the emission feature.

The other notable cluster of emission features (though individually less significant) is seen in the obscured observations at ~5-6 keV (A, Fig.~\ref{istogrammi}, bottom-left panel), is clearly visible in the corresponding RM (see Fig. \ref{pcfabsmaps}). This is detected to be more variable/sporadic in the 2000/2001 dataset (D). The nature of these features will be discussed in Sect.~\ref{bump}.
The U3 data were previously analyzed using the excess map technique by \cite{tombesi2007} with a different fitting procedure (described below). They report variability and modulation of a putative red and relativistically broadened wing of the Fe K line. In their scenario, the modulation was possibly due to the formation of spiraling arms within the accretion flow. It is worth noting, however, that the presence of this broad feature is not strongly supported by the analysis of the time averaged spectrum (R04). 
The recurrent, red-shifted feature reported in Tombesi et al. (2007) appears less prominent in our RM. This is probably due to the difference in the assumed baseline model (we assume a partial covering model for the continuum, rather than a totally covering one). Moreover, when fitting the continuum, we include a narrow Fe K$\alpha$ line in the model, rather than excluding the energy range where it contributes the most. More details on the comparison between our results and those reported in \cite{tombesi2007} are reported in Appendix \ref{appendice}.

\textbf{Features in absorption:} The most recurrent features measured in almost $\sim25\%$ of all the spectra are seen in absorption, between 6.7-6.9 keV (C and E in Fig. \ref{istogrammi} ). This absorption component is clearly visible also in the RM (see Fig. \ref{pcfabsmaps}). This is in agreement with results of R04 and M17. We also confirm that the feature's energy shifted by $\sim$200 eV, from $\sim$6.7 to $\sim$6.9 keV, between the unobscured and obscured datasets (see Fig.\ref{istogrammi}). Taking into account the calibration problems highlighted in Sect.~\ref{fekalphaline}, the net energy shift of the absorption line is more likely of $\sim$150 eV. The absorption line at ~6.7 keV is observed throughout all the unobscured observations, which span a period of about one year. Nonetheless, R04 reported a change in the EW between U2 and U3. However, it is worth noting that their analysis is most probably more sensitive to slight variations of the average values, while our analysis is more focused on searching variability on shorter time scales.  
Interestingly, we confirm the absence of UFOs in this source \citep{tombesi2010}. The results of the blind search in the energy band 8-10 keV are, in fact, consistent with what expected for pure casual events (see Table \ref{tab_sign}). 


\subsection{Possible Fe K$\alpha$ broadening during O1 and O2}
\label{bump}

To explain the origin of the  group of emission features detected at $\geq4\sigma$ in the obscured epoch in the 5-6 keV energy range (bottom-left panel of Fig.\ref{istogrammi}) and clearly visible in the corresponding RM (last two panels of Fig.\ref{pcfabsmaps}), we propose two different scenarios.\\ 
The first one ascribes it to the presence of the absorber. It may be an effect of the oversimplification of the fitted continuum model, meaning that ignoring the absorption lines and absorption edges due to the presence of an ionized absorber alters the whole continuum, occasionally resulting in excess emission redward of the Fe K$\alpha$ line. Another possibility in the same scenario is that the bump is produced by variations in the partial covering fraction happening on times shorter then our time resolution \citep{iso16}. This scenario is furthermore supported by the fact that in the unobscured epoch, where the absorber has a lower  column density  and is less ionized (R04) the features are detected at a lower significance and at lower energies (top left panel  of Fig.\ref{istogrammi}).\\
The other scenario is that these features are real, and associated with a relativistic wing of the neutral Fe K$\alpha$ line. To explore this option, we tested a relativistic model on the emission line detected at ~6 keV in the 12th 5ks spectrum O1, which we use as a test because of its significant broadening of $\sigma=260^{+130}_{-120}$ eV (Fig.\ref{kalfa_en_free_width}). We added a \texttt{laor} component \citep{laor91} to our baseline model, that already accounts for the narrow Fe K$\alpha$. The outer radius and emissivity index are unconstrained, so they are fixed to their default values of 400$\mathrm{R_g}$ and 3, respectively. The excess is well reproduced ($\Delta \chi^2=18.3$ for 3 $\Delta$d.o.f.) by a signal emitted from $\mathrm{R_{in}=99\pm52 \ R_{g}}$. At such distance, the Keplerian velocity is $\sim30000$ km/s, corresponding to a FWHM of $\sim0.64$ keV for the neutral Fe K$\alpha$ line at 6.4 keV. Thus, we do not expect to see much significant emission associated with it below $\sim6$ keV, were we indeed detect the majority of emission lines (group D, bottom-left panel Fig.~\ref{istogrammi}). If those features were real, they could be emitted from the base of the obscuring wind, as it could be expected following the models in \cite{deh20}.

\subsection{Fe K$\alpha$ modulation in O2}
\label{modulation}
\begin{figure}
	\centering
	\includegraphics[width=\hsize]{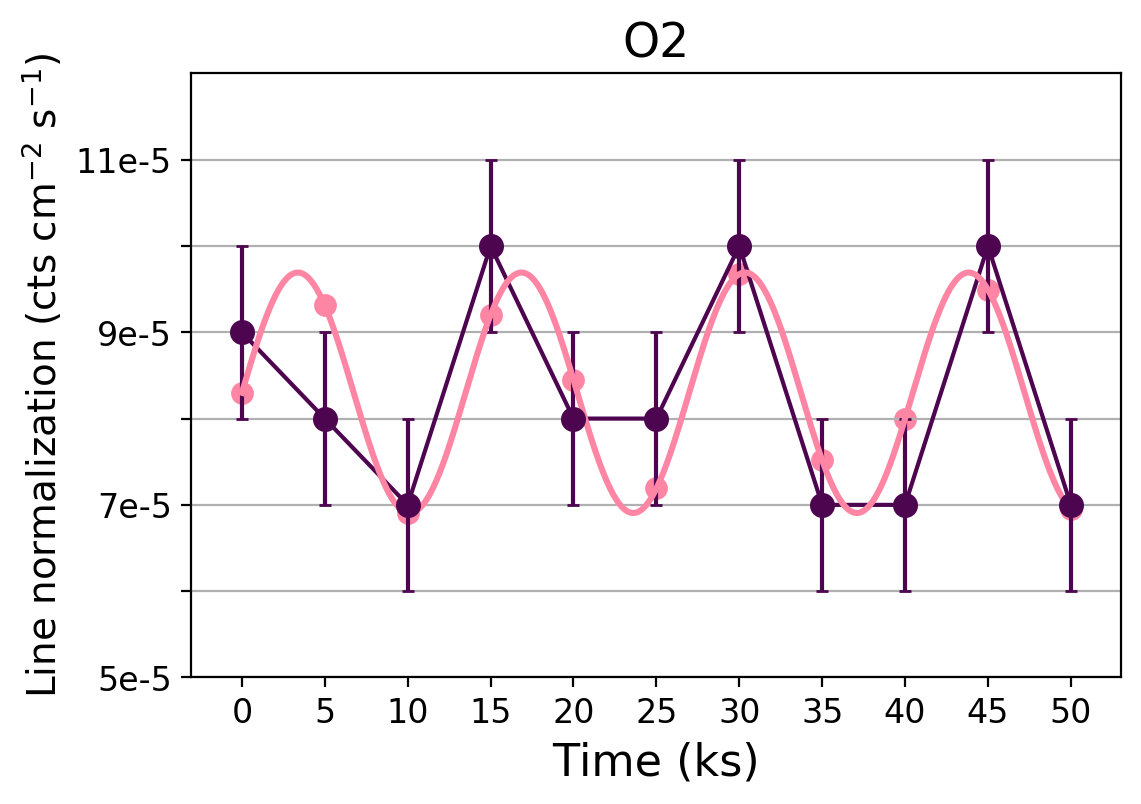}\\
	\caption{Tentative fitting of a sinusoid to the variations of the normalization of the Fe K$\alpha$.
	}
	\label{normalsin}
\end{figure}

In the residual maps of the latest observation (O2) (Fig.\ref{pcfabslinelinemaps}) there are four  quite evident peaks in the principal Fe K$\alpha$ line component. It is to be noted that their presence does not show a correlation with the $4-10$ keV light curve (last panel in Fig. \ref{lightcurves}), so the line is not simply following an intrinsic continuum variation.
The normalization of the narrow Fe K$\alpha$ line, as obtained from time-resolved spectral fitting of O2, is plotted in Fig.~\ref{normalsin}. This is tentatively fit with a sinusoidal function. 
The fit yields $\chi^2=5.3$ for 8 dof (10 data-points and 2 free parameters, period and normalization). This corresponds to a significance of $90\%$, thus not supporting the detection of a modulation. However, it is worth noting that even if a modulation is indeed present, the small number of sampled variability cycles ($\sim$4) would significantly reduce the significance of the signal.

The sinusoidal fit yields a best-fit value of $13\pm8$ ks for the putative period. This would correspond to a Keplerian orbit at $\sim$6 Rg. At such small distances we would expect prominent relativistic effects on the Fe K$\alpha$ line, which are not observed. Therefore, we conclude that the observed variations are most likely not associated with a modulation of the Fe K$\alpha$.

\begin{figure}
	\centering
	\includegraphics[width=\hsize]{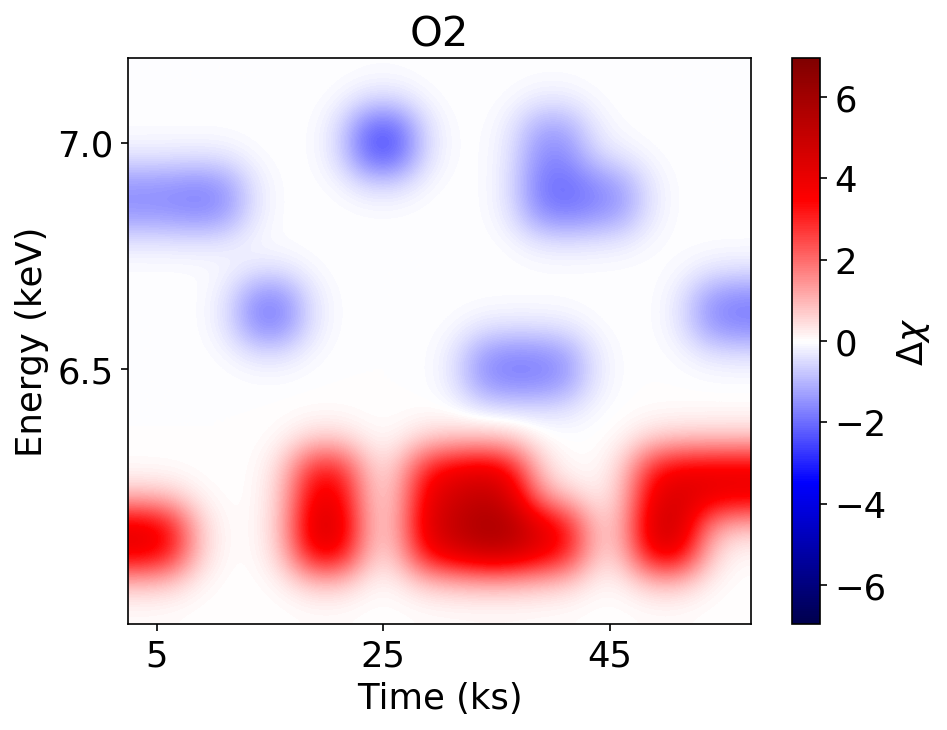}\\
	\caption{Residual map of O2 where we set a higher threshold in terms of $\Delta\chi$ to qualitatively highlight the possible link between the peaks of the Fe K$\alpha$ emission line (in red) and the most intense absorption dips (in blue).
	}
	\label{corrmap}
\end{figure}

Interestingly, the intensity of the narrow Fe K$\alpha$ line seems to decrease simultaneously with the appearance of the absorption features at ~6.7-6.9 keV, as highlighted in Fig.~\ref{corrmap}.
This suggests a relation between the two. In M17 the absorption features are ascribed to a highly ionized ($\log \xi\cong3.8$) clumpy medium outflowing at few thousand of km s$^{-1}$ in the broad line region. This high-ionization component is likely associated and spatially coexistent with the obscurer producing the eclipsing event, which has a lower ionization parameter of $\log \xi\cong1.8$. A gas with this value of $\xi$ could actually produce resonant absorption lines at energies consistent with the Fe K$\alpha$ for ionized species from Fe \textsc{x} up to Fe \textsc{xx} \citep{kallman2004}. In this case, the dips observed in the emission line would correspond to a partial absorption at that energy. This explanation is similar to that reported in M17 for the disappearance of the Fe K$\beta$ emission line; the main difference is that in the case of the Fe K$\alpha$ line the emission feature is not completely suppressed because of its intrinsically higher normalization.
In this scenario we can use the duration of the dips (and the absorption features at higher energy) to constrain the size of the clumps of the obscurer. If each clump has a large opacity, we can assume an "on/off" effect due to their passage through the line of sight. Using the distance and velocity values found in M17 for the obscurer, $\sim7 - 10$ light days from the X-ray source (few $\mathrm{10^3\ R_g}$), with an orbital velocity of $\mathrm{\sim3500 - 4200 \ km \ s^{-1}}$, and considering the duration of the dips to be from 5ks to 10 ks, we can estimate the clumps extent to be in the range $\sim1.7-4.2\times 10^{13}$cm.

While the peaks in O2 are the most evident, Fig.~\ref{pcfabslinelinemaps} shows that the intensity of the neutral Fe K$\alpha$ emission line varies also in other observations.
For O1 we can assume that the lower ionization component is the same as in O2 (M17), and since the Fe K$\alpha$ line dips have the same duration of 5-10 ks we obtain the same results on the clumps sizes.

In the unobscured epoch, R04 reports the presence of three different absorbing components at different ionization levels. Among them, the medium with N$\mathrm{_H\cong4.4\times 10^{22}\ cm^{-2}\ and\ log \xi\cong3}$ can possibly absorb part of the neutral Fe K$\alpha$ emission line. The entity of this absorption is actually consistent with the variations we measure in the normalization of the line (a few percents of its value), it is to be noted tough that this quantity is also of the same order of magnitude of the error of the parameter. However if we hypothesize the variations to be real, we measure dips with a duration between 5 ks and 20 ks. With a distance of $\sim2\times10^{17}$cm (R04), assuming a keplerian orbit we find the size of the clouds in the range $0.7-2.8\times10^{11}$cm. 
The dimensions of the clouds in both epochs are consistent with those found in the BLR of NGC 1365 by \cite{risaliti09}.

\section{Conclusions}
In this paper we presented a time-resolved spectral analysis of all \xmm\ observations (exposure $\geq$10 ks) of NGC 3783. We first performed a blind search to detect any significant feature either in emission or in absorption and then produced residual maps to trace their evolution over time and energy, searching for variable structures that might be indicative of motions or changes in the physical state of the accreting/outflowing gas near the BH. 
Running the blind search and then producing the residual maps, we examined both positive and negative deviations found in the 4-10 keV energy band. Our approach slightly differs from past works, as those work mostly focused on studying only excess emission features  and not absorption ones (e.g. \citealt{turner2006,tombesi2007,demarco2009,nardini2016,marinucci2020}). \\
This study is complementary to the work presented in DM20, who focused on the study of the soft X-ray band variability properties of the source, and to \citet{tombesi2007}, who characterized the variability of emission features in the unobscured dataset.

In this work we found that NGC 3783 5ks spectra show discrete features detected at $\geq90\%$ (Fig. \ref{istogrammi}) at different times both in emission and in absorption, and the distributions in energy of these features change between the datasets from the two different states. In particular all detections show a higher significance during the obscured phase than in the unobscured one and are not correlated with variations in flux. 
Furthermore we observed that some of the lines appear multiple times at the same energies, reaching a global significance higher than 3$\sigma$:
\begin{itemize}
    \item in the 4-6 keV energy band we detect clusters of emission features in both our datasets, with a higher significance in the obscured one, whose presence we link to complex effects of the absorption mediums; 
    \item the spectra from the unobscured epoch present an emission feature ascribable to a blend of Fe neutral K$\beta$ and ionized Fe K$\alpha$;
    \item absorption features are detected at $\sim6.6$ keV in the unobscured dataset and at $\sim6.7-6.9$ in the obscured one, produced by two distinct absorbers.
\end{itemize}
In the residual maps we did not find any particular pattern or evolution in most of the data, except for O2 where we identified a possible modulation in the neutral Fe K$\alpha$ signal that seems to be linked to variations in the recurrent absorption feature at slightly higher energies: we interpreted it as an effect of the clumpiness of the absorber medium and inferred the dimensions of the clumps.

The described analysis and the methodology used in this work can be extended to further bright Seyfert~1 galaxies with long and multiple \xmm\ observations, and have strong potentialities for next-generation X-ray instruments, as those onboard of the ESA mission {\it Athena} \citep{nandra13}.

\begin{acknowledgements}
We thank the anonymous referee for her/his constructive comments.
This work is based on observations obtained with XMM-Newton, an ESA science mission with instruments and contributions directly funded by ESA Member States and NASA. Partial support to this work was provided by the European Union Horizon 2020 Programme under the AHEAD2020 project (grant agreement number 871158). We acknowledge financial support from ASI under grants ASI-INAF I/037/12/0 and n. 2017-14-H.O, and from the grant PRIN MIUR contract no. 2017PH3WAT ("Black Hole winds and the Baryon Life Cycle of Galaxies: the stone-guest at the galaxy evolution supper"). DC thanks AP and TS. BDM acknowledges support via Ram\'on y Cajal Fellowship RYC2018-025950-I. P.O.P. acknowledges financial support from the High Energy National Programme (PNHE) of CNRS and from the french spatial agency CNES. E.B. was supported by a Center of Excellence of The Israel Science Foundation (grant No. 2752/19). SRON is supported financially by NWO, the Netherlands Organization for Scientific Research.
\end{acknowledgements}

\bibliographystyle{aa} 
\bibliography{bib.bib}{}
\begin{appendix}
\section{Continuum fitting in residual maps}
\label{appendice}

As stated in Sect.~\ref{discussion}, our residual maps have been produced in a slightly different fashion that those described by \cite{tombesi2007}. Aside from the major step of employing all of the residuals, positive and negative, and not only the positive ones, we used a different fitting procedure.
 
\begin{figure}
	\centering
	\resizebox{\hsize}{!}{
	\includegraphics{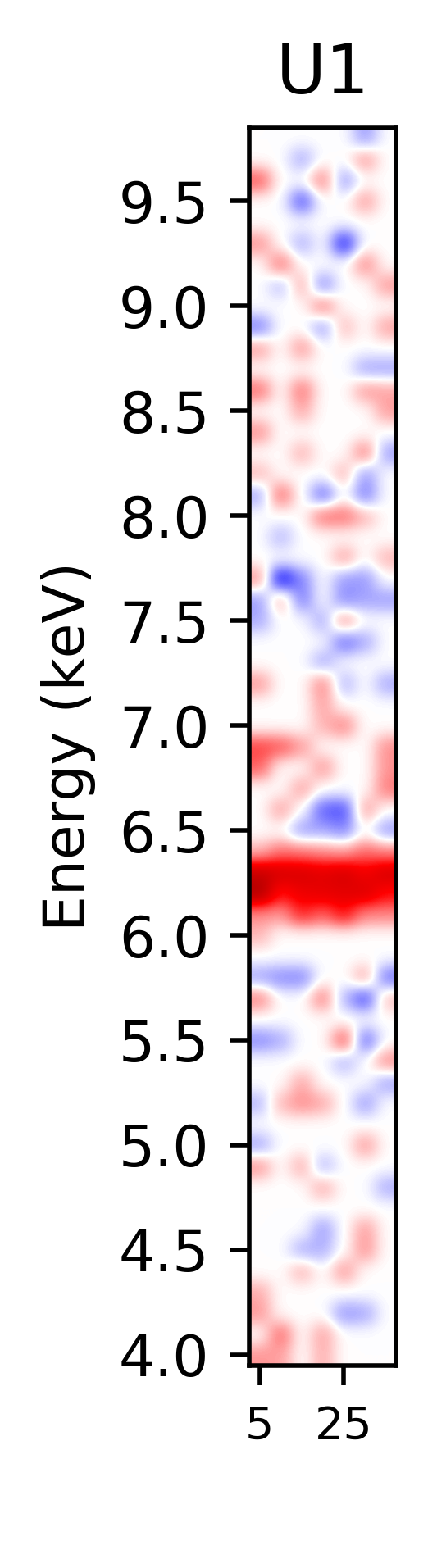}
	\includegraphics{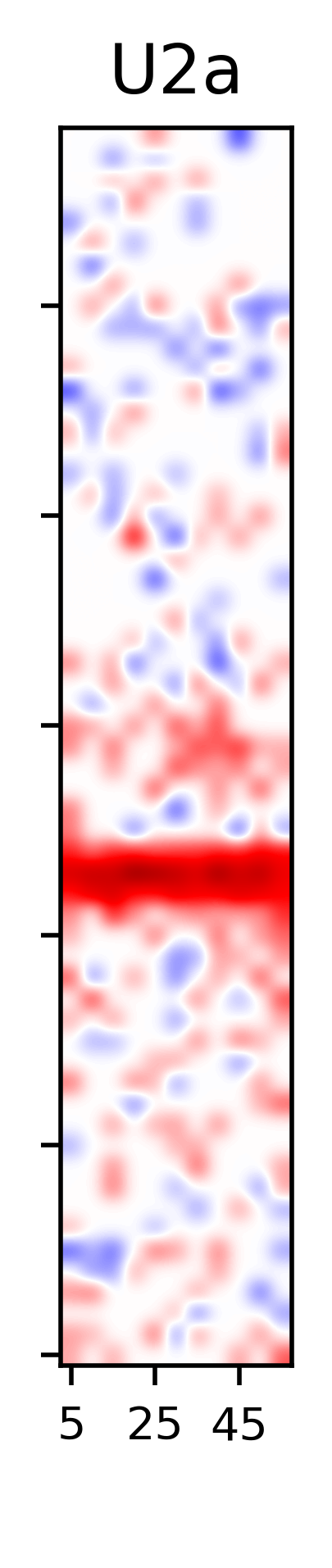}
	\includegraphics{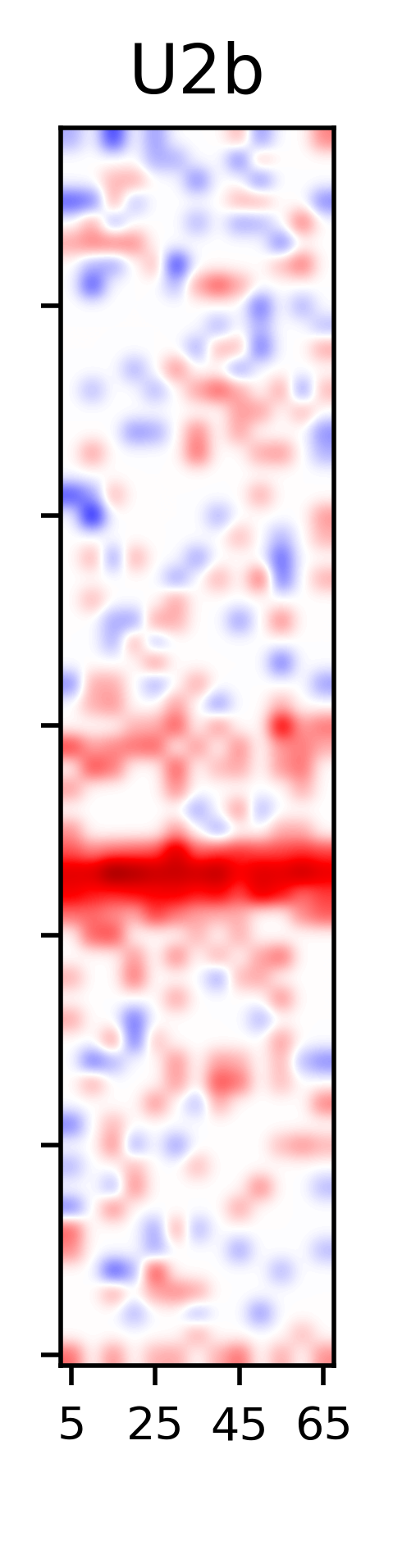}
	\includegraphics{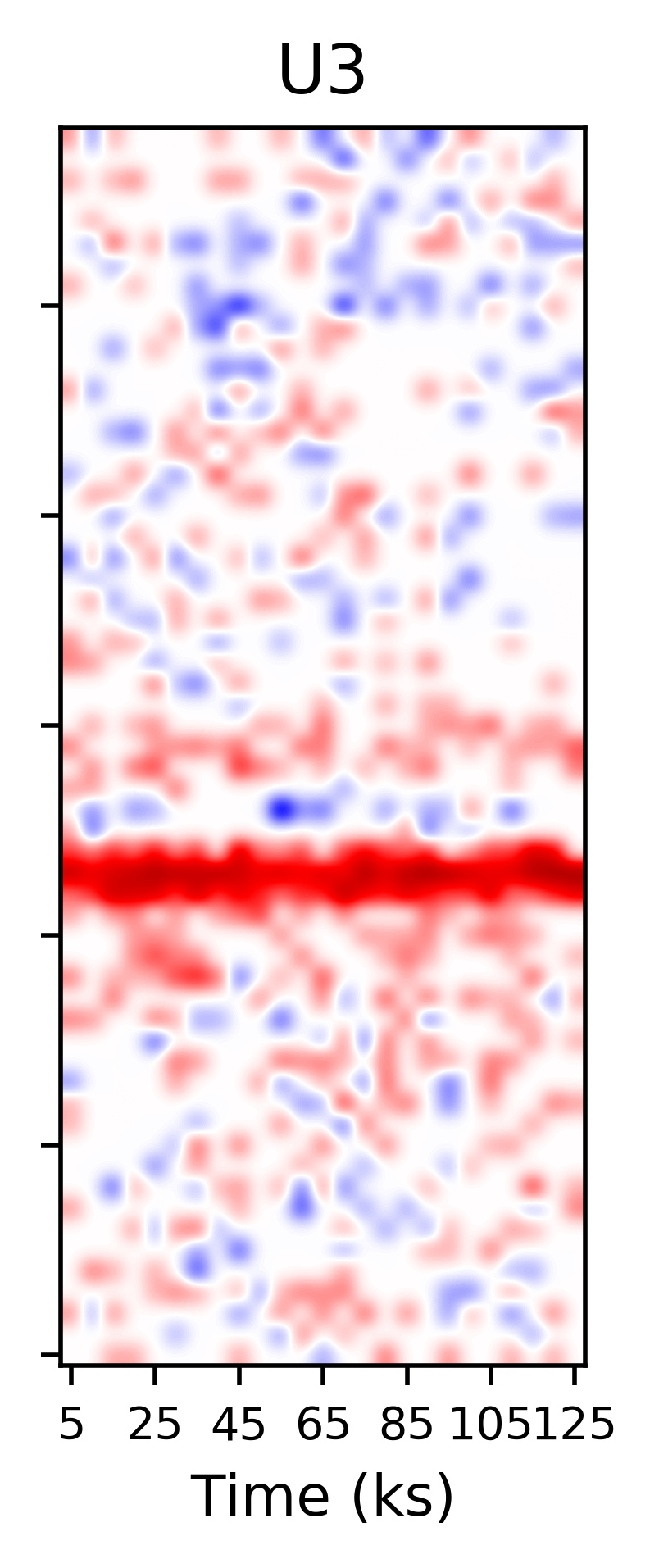}
	\includegraphics{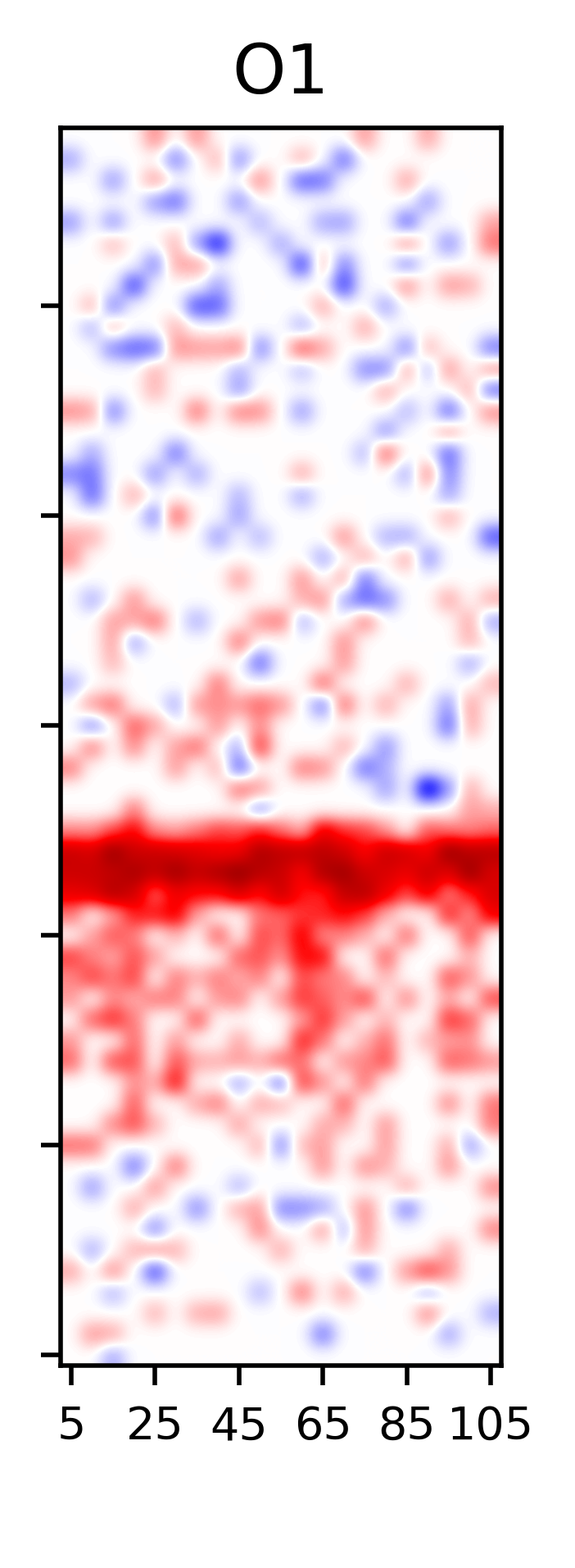}
	\includegraphics{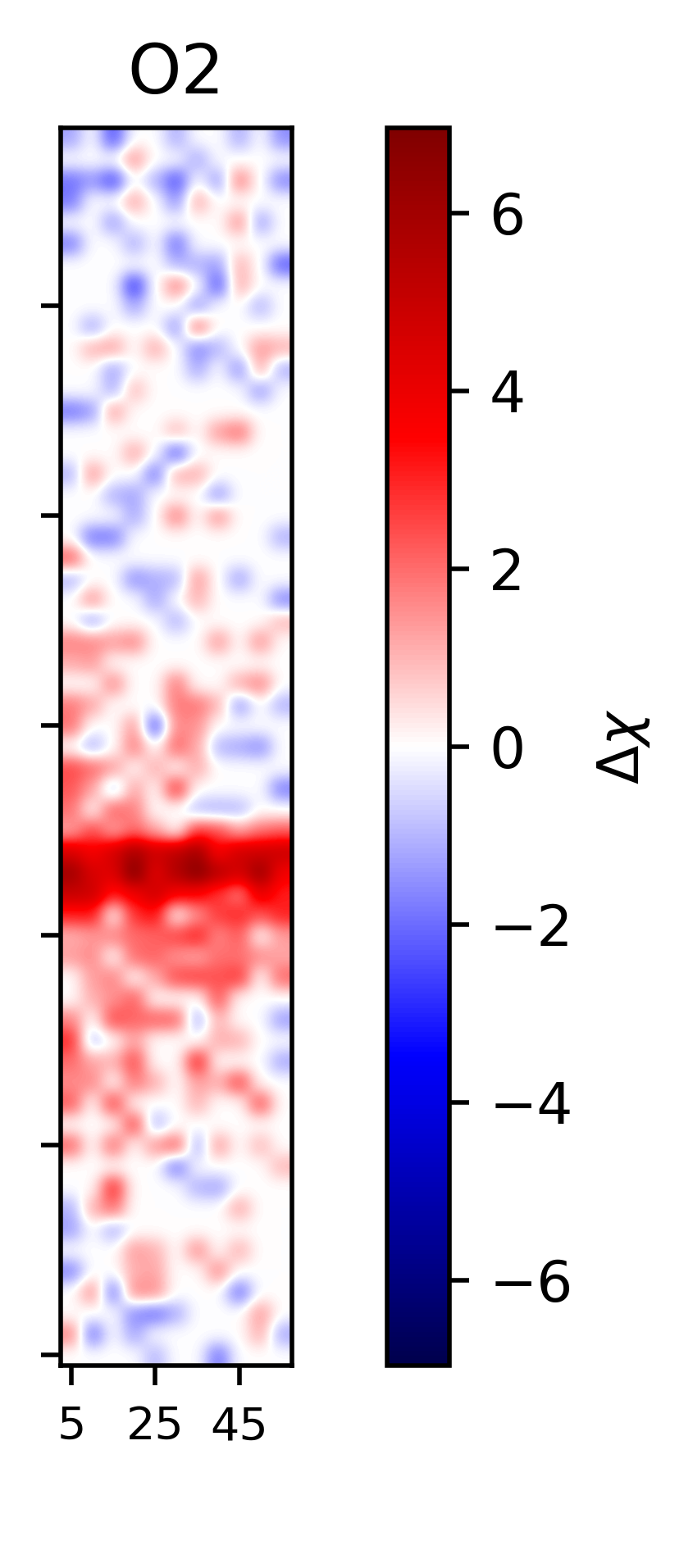}}\\
		\resizebox{\hsize}{!}{
	\includegraphics{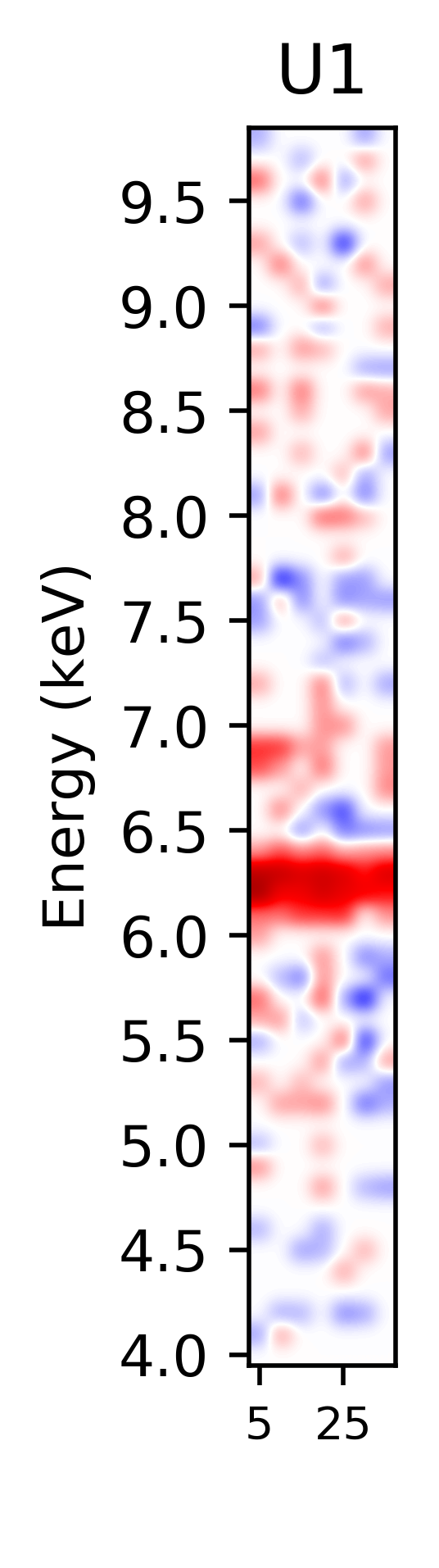}
	\includegraphics{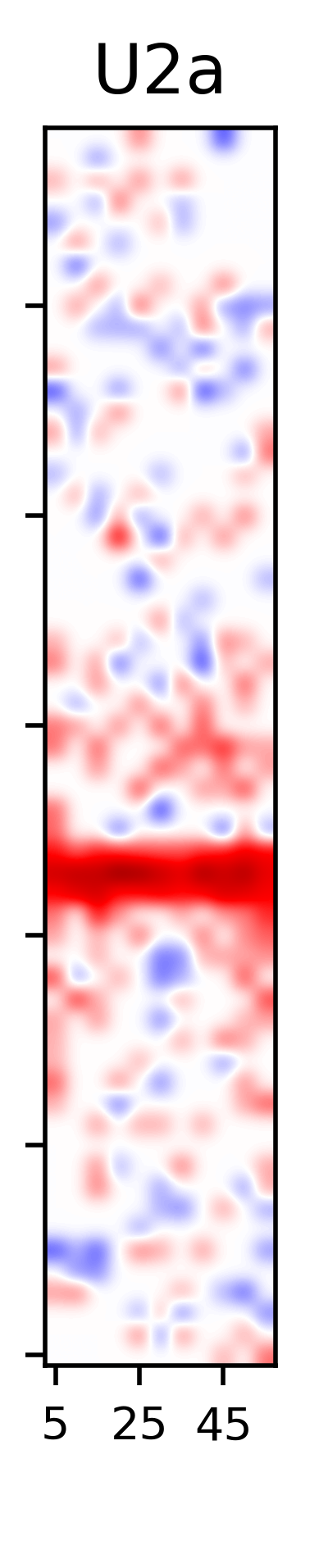}
	\includegraphics{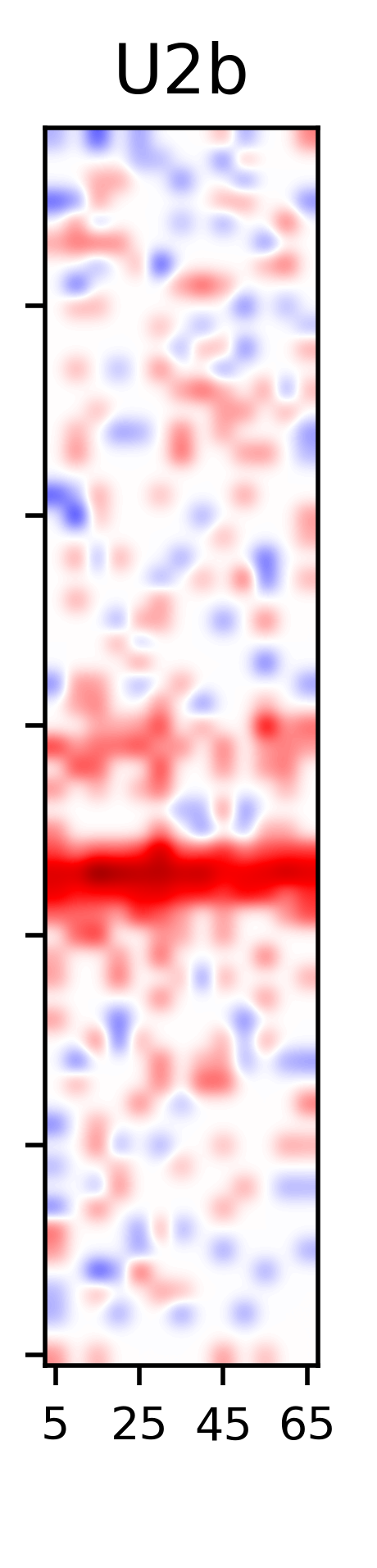}
	\includegraphics{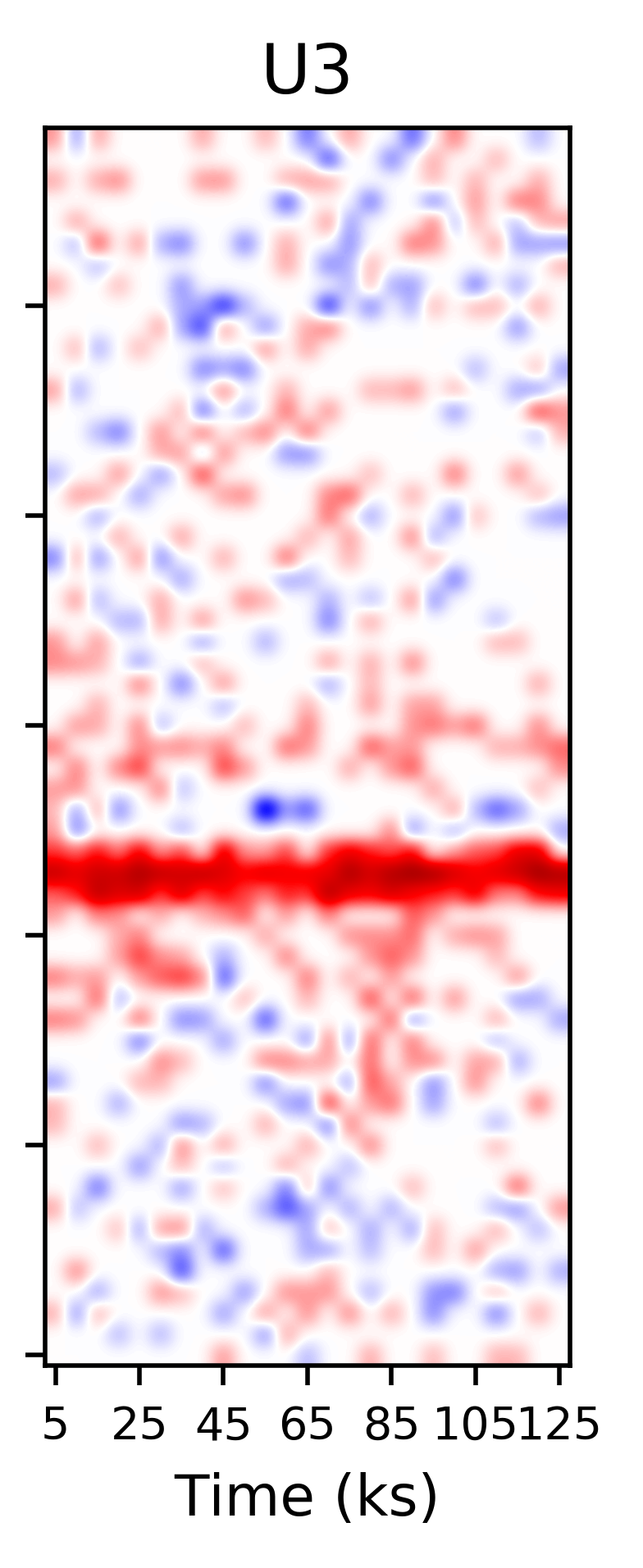}
	\includegraphics{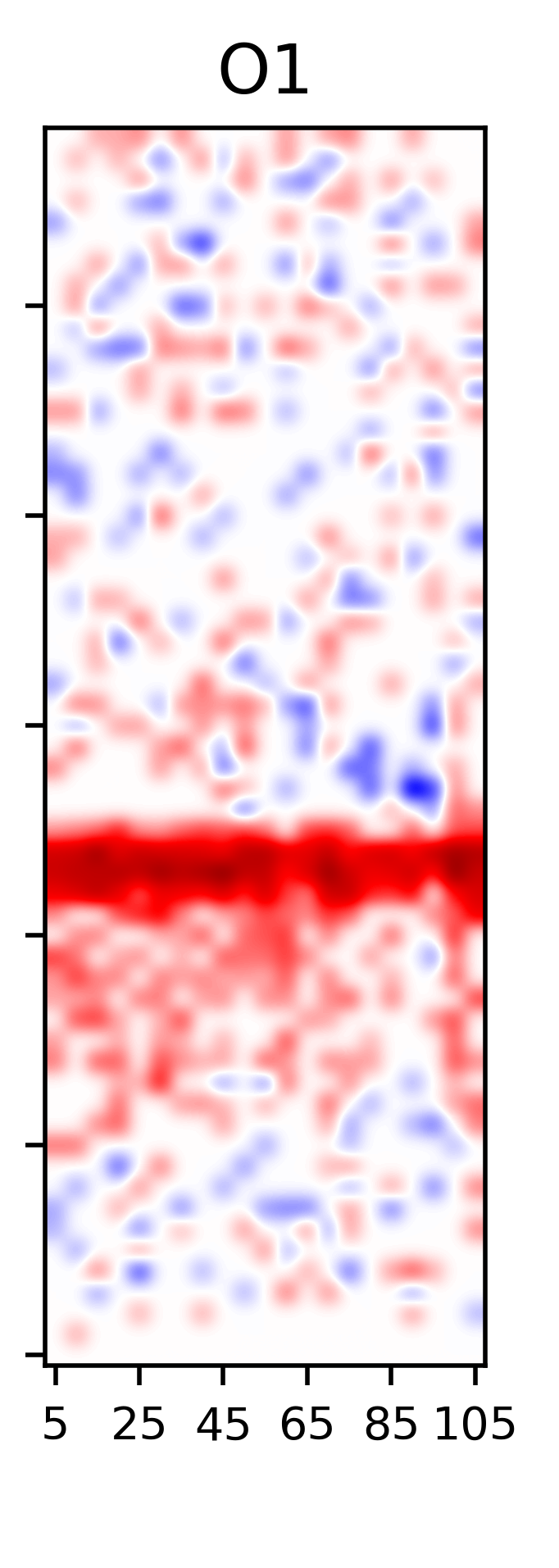}
	\includegraphics{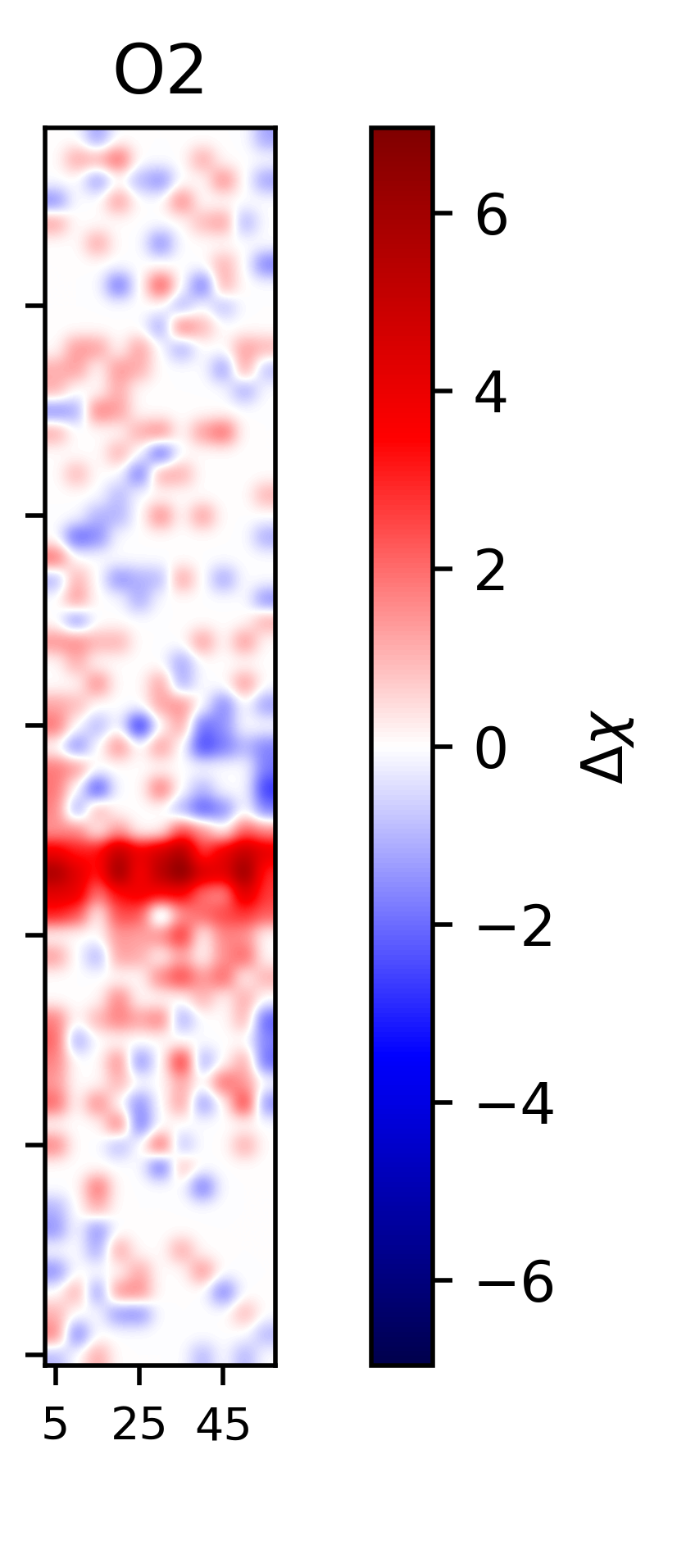}}
	\caption{RM produced using a power law plus total covering cold absorption fitted only in the 4-5 keV and 7-10 keV energy bands, following the indications of \cite{tombesi2007}. The value of the column density of the absorber is fixed to the value measured in average spectra in top panel maps and free to vary in the bottom panel ones.}
	\label{appendixmaps}
\end{figure}

To study the emission features, \cite{tombesi2007} modeled the continuum as a power law plus a cold absorption component fixed to be equal to the value measured in analyzing the average spectrum. Then they fitted the continuum considering the 4-5 keV and 7-9 keV band to avoid the Fe K$\alpha$ line. Applying the exact same procedure we produced the RM shown in the top panel of Fig.~\ref{appendixmaps}.

Since we were actually interested in following the variations of the absorptions features, produced indeed by the absorber, we produced another set of RM with the same fitting band and model, this time leaving the column density free to vary in each spectrum. They are displayed in the bottom panel of Fig.~\ref{appendixmaps}.

While these two sets of maps do not show significant divergences between each other, as indeed predicted by \cite{tombesi2007}, the difference among them and those we presented in Fig.~\ref{pcfabslinelinemaps} is quite evident. The absorption features we found at $\sim$6.6-6.9 keV are only marginally present in the RM in Fig.\ref{appendixmaps}. Instead, the excess at energies below the Fe K$\alpha$ line appear to be far more intense and, in particular, in U3 we find the strong feature discussed by \cite{tombesi2007}. It is to be noted that the normalization (colour) of the residual maps is the same for all sets.

\begin{figure*}
	\centering
	\resizebox{\hsize}{!}{
	\includegraphics{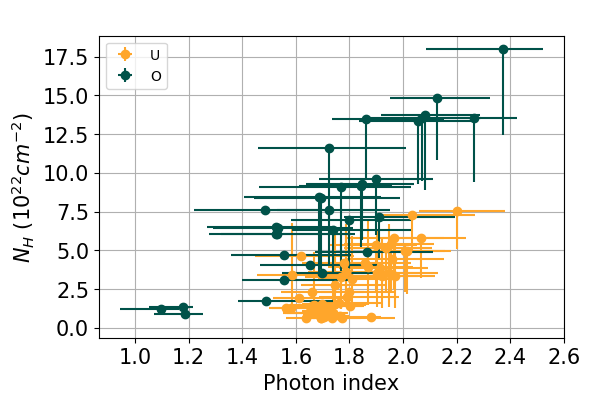}
	\includegraphics{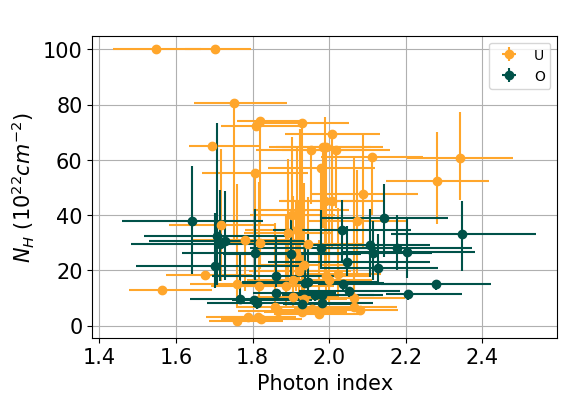}}
	\resizebox{\hsize}{!}{
	\includegraphics{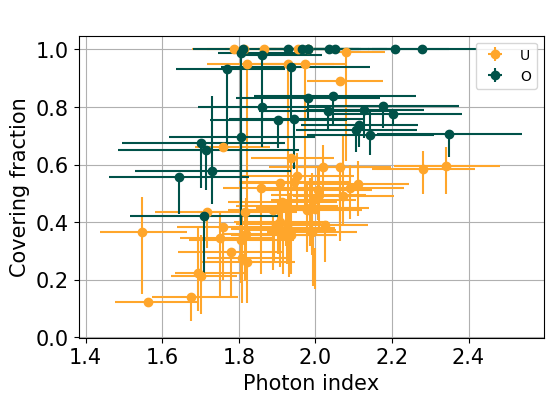}
	\includegraphics{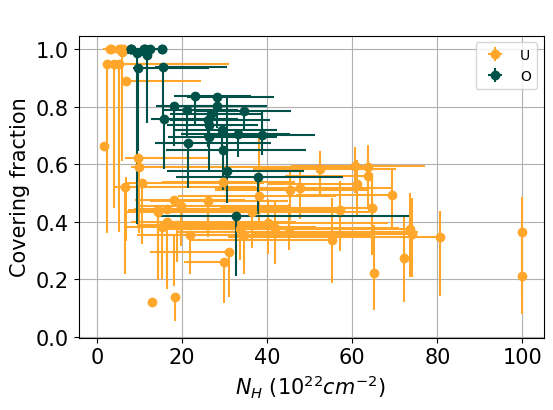}}
	\caption{Top left panel: column density of the total covering cold absorption vs power law photon index, fitted only in the 4-5 keV and 7-10 keV energy bands. The correlation among the values is assessed at $\gg5\sigma$ for both the unobscured and obscured epoch. Top right panel: column density of the partial covering cold absorption vs power law photon index, fitted in the 4-10 keV energy range. The degeneracy between the two parameters is not present in this case, with a correlation probability $\lesssim 20\%$. Bottom left panel: covering fraction of the partial covering absorber vs power law photon index. Bottom right panel: covering fraction vs column density of the absorber. Data relative to unobscured and obscured epoch are reported in yellow and green respectively.}
	\label{continuumparamters}
\end{figure*}

The first difference between the production of the maps shown in Fig.~\ref{pcfabsmaps} and those in Fig.\ref{appendixmaps} is the energy band used to fit the model. The use of the complete 4-10 keV range, without excluding the Fe K$\alpha$ band, allows for a better anchorage of the continuum, and this could be at the base of the divergence of the sets of maps. Then we analyzed the continuum parameters: in the top left panel of Fig.~\ref{continuumparamters} we plotted the best fit parameters for the absorber column density and the power law photon index. To quantify the relation among them, we calculated the Pearson Correlation Coefficient $\rho$ for the two epochs, and obtained $\mathrm{\rho_U=0.78}$ and $\mathrm{\rho_O=0.88}$, both corresponding to a correlation probability higher than 5$\sigma$. This degeneracy between these two parameters disappears when using the partial covering baseline model, described in Sect.~\ref{baselinemodel}. Its best fit parameters are plotted in the bottom panel of Fig.~\ref{continuumparamters}, and the correlation probability calculated for these data drops at 21$\%$ for the unobscured epoch ($\mathrm{\rho_U=-0.04}$) and 17$\%$ for the obscured one ($\mathrm{\rho_O=0.04}$).
On the one hand, it is worth considering that we are in any case dealing, by construction, with limited statistics. That means that also our approach is prone to introduce some systematics, as seen in the bottom panels of  Fig.~\ref{continuumparamters}. This is somehow inevitable given the experimental condition that we are using. On the other hand, we already knew that our baseline model provides a limited description of the complex absorption that is known to occur in NGC 3783 \citep{mehdipour2017,mao19}. Far from our scopes, we are not investigating here in detail the characteristics of the various ionized absorbers that have been previously reported in literature; we want to stress here is that at least part of excesses recorded between 5-6 keV are possibly explained by the simplified absorption scheme used in producing RM. A significant step forward in time-resolved spectral studies will be possible with X-ray telescopes with larger effective area, which would allow us to probe short timescales (i.e., by using short time slices) without losing energy resolution. Specifically, the X-IFU onboard Athena \citep{nandra13,barret18} in the foreaseable future will provide a major advance in the research field described in this paper.

\end{appendix}
\end{document}